\documentclass[prb,twocolumn,10pt,aps,showpacs,superscriptaddress,longbibliography,floatfix]{revtex4-1}

\usepackage{amsmath,amssymb,physics}
\usepackage{graphicx}
\usepackage{dcolumn}
\usepackage{bm}
\usepackage{color}
\usepackage{url}
\usepackage[colorlinks=true,linkcolor=blue,citecolor=blue,urlcolor=blue,breaklinks]{hyperref}

\usepackage{array}

\usepackage{soul}
\usepackage{mathtools}

\begin{document}

\title{Skyrmionic chains and lattices in $s+id$ superconductors}

\author{Ling-Feng Zhang}\email{lingfeng.zhang@uantwerpen.be}
\affiliation{Department of Physics, Shanghai University, Shanghai 200444, China}
\affiliation{Departement Fysica, Universiteit Antwerpen, Groenenborgerlaan 171, B-2020 Antwerpen, Belgium}
\author{Yan-Yan Zhang}
\affiliation{Department of Physics, Shanghai University, Shanghai 200444, China}
\author{Guo-Qiao Zha}
\affiliation{Department of Physics, Shanghai University, Shanghai 200444, China}
\author{M. V. Milo\v{s}evi\'{c}}
\affiliation{Departement Fysica, Universiteit Antwerpen, Groenenborgerlaan 171, B-2020 Antwerpen, Belgium}
\author{Shi-Ping Zhou}\email{spzhou@shu.edu.cn}
\affiliation{Department of Physics, Shanghai University, Shanghai 200444, China}

\begin{abstract}
We report characteristic vortex configurations in $s+id$ superconductors with time reversal symmetry breaking, exposed to magnetic field. A vortex in the $s+id$ state tends to have an opposite phase winding between $s-$ and $d-$wave condensates. We find that this peculiar feature together with the competition between $s-$ and $d-$wave symmetry results in three distinct classes of vortical configurations. When either $s-$ or $d-$ condensate absolutely dominates, vortices form a conventional lattice.  However, when one condensate is relatively dominant, vortices organize in chains that exhibit skyrmionic character, separating the chiral components of the $s \pm id$ order parameter into domains within and outside the chain.  Such skyrmionic chains are found stable even at high magnetic field. When $s-$ and $d-$ condensates have a comparable strength, vortices split cores in two chiral components to form full-fledged skyrmions, i.e. coreless topological structures with an integer topological charge, organized in a lattice. We provide characteristic magnetic field distributions of all states, enabling their identification in e.g. scanning Hall probe and scanning SQUID experiments. These unique vortex states are relevant for high-T$_c$ cuprate and iron-based superconductors, where the relative strength of competing pairing symmetries is expected to be tuned by temperature and/or doping level, and can help distinguish $s+is$ and $s+id$ superconducting phases.
\end{abstract}


\date{\today}

\maketitle

\section{Introduction}\label{sec:1}

Over the years there has been large experimental and theoretical interest in superconducting states that exhibit time-reversal symmetry breaking (TRSB).  Such states can appear in superconductors with two or more coupled condensates when the phase differences between condensates are neither $0$ nor $\pi$ in the ground state, and often host novel features.\cite{milosevicEmergentPhenomenaMulticomponent2015}  For example, the TRSB $p_x + ip_y$ and $d_{x^2-y^2}+ id_{xy}$ states are topologically non-trivial, allowing for characteristic edge states and hosting Majorana zero modes for topological quantum computing.\cite{satoTopologicalSuperconductorsReview2017}  The $p_x + ip_y$ state has been proposed for layered ruthenate superconductor Sr$_2$RuO$_4$,\cite{mackenzieSuperconductivityMathrmSrMathrmRuO2003, kallinChiralPwaveOrder2012} and the $d_{x^2-y^2}+ id_{xy}$ state for hexagonal systems in e.g. doped graphene \cite{nandkishoreChiralSuperconductivityRepulsive2012, black-schafferChiralWaveSuperconductivity2014} and SrPtAs.\cite{fischerChiralWaveSuperconductivity2014}

Superconductors with TRSB often exhibit rich and novel phenomena associated with topological defects.\cite{tanakaMulticomponentSuperconductivityBased2015, linGroundStateCollective2014} Being multi-component as a rule, such superconductors can support non-monotonic interactions between vortices, leading to nonuniform vortex patterns.\cite{babaevSemiMeissnerStateNeither2005, linVortexStatesPhase2011, xuSimulationPhaseDiagram2014} Leggett mode and phase solitons can arise as topological defects associated with non-trivial phase difference between different components.\cite{tanakaSolitonTwoBandSuperconductor2001, linPhaseSolitonsMultiband2012} Furthermore, in the state with TRSB, domain walls may separate areas with different TRSB ground states.\cite{matsumotoQuasiparticleStatesSurface1999, garaudDomainWallsTheir2014, zhangTopologicalPhaseTransitions2017} A closed domain wall of such kind with attached vortices exhibits non-zero, integer topological charge, defining such a topological defect as a skyrmion.\cite{babaevHiddenSymmetryKnot2002, zhang_electronic_2016}  Skyrmions have been studied to date in $p+ip$ state,\cite{zhang_electronic_2016, garaudLatticesDoublequantaVortices2016, garaudPropertiesSkyrmionsMultiquanta2015, fernandezbecerraVorticalSkyrmionicStates2016, fernandezbecerraDynamicsSkyrmionsEdge2017} $s+is$ state,\cite{garaudChiralMathbbCPSkyrmions2013} and in nematic superconductors.\cite{zyuzinNematicSkyrmionsOddParity2017} Such skyrmion was first identified as a coreless vortex in superfluid $^3$He-$A$.\cite{hoCorelessVorticesSuperfluid1978}  It is also often referred to as a Mermin-Ho\cite{merminCirculationAngularMomentum1976,  mizushimaMerminHoVortexFerromagnetic2002} or Anderson-Toulouse vortices\cite{andersonPhaseSlippageVortex1977} in Bose-Einstein condensates.\cite{PhysRevLett.81.742, PhysRevLett.90.140403, PhysRevA.81.033629, mizushimaCorelessSingularVortex2004, choiObservationTopologicallyStable2012, orlovaSkyrmionicVortexLattices2016, PhysRevA.91.043605, PhysRevA.77.033621, PhysRevLett.112.075301, alkhawajaSkyrmionsFerromagneticBose2001a}

Over the last decade, the immense interest in iron-based superconductors has stimulated extensive studies on the $s+is$ and $s+id$ superconducting phases with TRSB.\cite{kamiharaIronBasedLayeredSuperconductor2006, kamiharaIronBasedLayeredSuperconductor2008, linDistinguishingIdPairing2016, garaudMicroscopicallyDerivedMulticomponent2017} The majority of researchers believe that most moderately-doped iron-based superconductors possess an $s_\pm$ pairing symmetry due to strong interaction between electron and hole Fermi pockets.\cite{mazinSuperconductivityGetsIron2010, kurokiUnconventionalPairingOriginating2008}  For example, the Ba$_{1-x}$K$_x$Fe$_2$As$_2$ favors the $s_\pm$ state at $x=0.4$, where the superconducting gap changes sign between electron and hole pockets, consistent with ARPES,\cite{dingObservationFermisurfaceDependent2008} thermal conductivity\cite{luoQuasiparticleHeatTransport2009} and neutron scattering experiments.\cite{christiansonUnconventionalSuperconductivityBa02008}  However, further potassium doping results in topological changes in the Fermi surface, where the electron pockets disappear and additional hole pockets appear.\cite{xuPossibleNodalSuperconducting2013}  Theorists have suggested another $s_\pm$ pairing or a $d$-wave pairing in that case.  Therefore, one expects an intermediate $s+is$ \cite{maitiStateBrokenTimereversal2013, maitiSpontaneousCurrentsSuperconductor2015} or $s+id$ state \cite{leePairingStateTimeReversal2009} at some doping level between $x=0.4$ and $1$, and the competition between the two different pairing symmetries depends on the doping level.  Recently, a TRSB state was found in the Ba$_{1-x}$K$_x$Fe$_2$As$_2$ at $x \approx 0.75$,\cite{grinenkoSuperconductivityBrokenTimereversal2017} which is consistent with the theoretical argument.

To distinguish between $s+is$ and $s+id$ states in real materials, great efforts have been made on understanding their superconducting properties.  Spontaneous magnetization near an impurity site has been studied in detail for both states.\cite{leePairingStateTimeReversal2009, maitiSpontaneousCurrentsSuperconductor2015, linDistinguishingIdPairing2016}  Thermoelectric effect was found to be unconventional in the $s+is$ state.\cite{silaevUnconventionalThermoelectricEffect2015, garaudThermoelectricSignaturesTimeReversal2016}  In particular, topological defects such as vortices, domain walls, and skyrmions have been studied for the $s+is$ state under magnetic field.\cite{garaudDomainWallsTheir2014}  The corresponding study for the $s+id$ state is lacking, and that is the primary objective of this paper.

To date, only single vortex and vortex lattice solutions have been studied in the mixed $s$- and $d$-wave superconducting states, i.e. $s+d$ state\cite{renGinzburgLandauEquationsVortex1995, xuStructuresSingleVortex1996, franzVortexStateDwave1996} and $s+id$ state.\cite{liVortexStructureWave1999}  Based on symmetry considerations, when the $d$-wave component is dominant, the $s$-wave component is known to have an opposite phase winding relative to the $d$-wave component near a vortex in the $s+d$ state.\cite{volovikSuperconductivityLinesGAP1993}  As we will show, partially due to the latter feature, the TRSB $s+id$ state under magnetic field can exhibit rich and unique phenomena associated with vortex matter, domain walls and skyrmions. In particular, we study the vortical configurations under different levels of competition between participating condensates.  We find that when the $s+id$ state is changed from a $d$-wave dominant state to an $s$-wave dominant state (induced by e.g. changing temperature or doping level in iron-based superconductors), the system can host three distinct classes of vortex matter, as seen from Fig.~\ref{fig.gs}, namely the vortex lattice, skyrmionic vortex chains and skyrmionic lattices. For each of the classes their characteristic magnetic field distributions enable experimental identification, in e.g. scanning Hall probe and scanning SQUID microscopy.  In particular, we identify that the competing orders between different symmetries in the $s+id$ state result in the skyrmionic vortex chain that has no analogy in the $p+ip$ and $s+is$ state, being important for distinguishing the $s+id$ state from the $s+is$ state in unconventional superconductors.  

The paper is organized as follows. In Section \ref{sec:2} we introduce the mixed $s$- and $d$-wave Ginzburg-Landau formalism that is used in the simulations of vortex configurations.  In section \ref{sec:3}, we discuss the ground-state properties as the competition between two condensates is varied.  Then, in sections \ref{sec:3.1}, \ref{sec:3.2} and \ref{sec:3.3}, we detail the results with respect to the three distinct classes of vortex matter, found as the result of the competition of pairing symmetries. The characteristics of the latter states are then compared to those found in the $s+is$ state in section \ref{sec:4}.  Our conclusions are summoned in section \ref{sec:5}.

\section{Theoretical formalism}\label{sec:2}

The dimensionless Ginzburg-Landau (GL) free energy functional describing the mixed $s$- and $d_{x^{2}-y^{2}}$-wave symmetry with two order parameters $(\psi_d, \psi_s)$ reads:\cite{renGinzburgLandauEquationsVortex1995}
\begin{align}\label{eq.GL}
F &= \frac{1}{\Omega} \int  \bigg\{-2\alpha_s|\psi_s|^2-|\psi_d|^2+\frac{4}{3}|\psi_s|^4+\frac{1}{2}|\psi_d|^4 \notag\\
&+2|\Pi\psi_s^{\ast}|^2+|\Pi\psi_d^\ast|^2 +\kappa^2(\nabla\times \vec{A})^2\notag\\
&+\underbrace{\Pi{_{x}^{\ast}}\psi_s\Pi_x\psi_d^{\ast}-\Pi_y^{\ast}\psi_s\Pi_y\psi_d^{\ast} + c.c.}_{\zeta} \notag\\
& +\underbrace{\frac{8}{3}|\psi_s|^2|\psi_d|^2}_{\eta} +\underbrace{\frac{2}{3}(\psi_s^{\ast2}\psi_d^2+c.c.)}_{\delta} \bigg\} d\Omega,
\end{align}
where $\Pi\equiv i\nabla-\vec{A}$ is the momentum operator, $\Omega$ is specimen volume, and the $\zeta$, $\eta$ and $\delta$ terms are direct density coupling, mixed gradient and Josephson coupling, respectively.  The $\delta$ term is associated with the relative phase between both condensates, and can be rewritten as 
\begin{equation}
    \frac{2}{3}(\psi_s^{\ast2}\psi_d^2+c.c.) = \frac{4}{3}|\psi_s|^2|\psi_d|^2 \mathrm{cos}(2\theta_{sd}),
\end{equation}
with relative phase $\theta_{sd}=\theta_s-\theta_d$.  The mixed gradient $\zeta$ term is not symmetric with respect to $x$ and $y$ direction, which plays a key role in the resulting vortex structure.

Eq.~\eqref{eq.GL} contains only two adjustable parameters, namely the relative nominal strength of the two condensates $\alpha_s$ in absence of magnetic field, and the GL parameter $\kappa$ that controls the magnetic screening of the applied field. $\alpha_s$ may be expressed as a function of temperature $T$ as $\alpha_s = \ln(T_s/T)/\ln(T_d/T)$, where $T_s$ and $T_d$ are the nominal critical temperatures of the $s$-wave and the $d$-wave superconducting order, respectively, with $T_s \propto e^{-1/(N(0)V_s)}$ and $T_d \propto e^{-1/(N(0)V_d)}$ (where $N(0)$ is the density of states at the Fermi surface, and $V_s$ and $V_d$ are the effective attractive interaction strengths in the $s$- and $d$-wave channels, respectively).  Therefore, $\alpha_s$ determines the relative strength between the condensates, i.e. $|\psi_d|/|\psi_s|$, which can be varied by temperature $T$ and/or $V_s$ and $V_d$ through variation of the doping level. For $T_d>T_s$, $\alpha_s<1$.

The remaining coefficients in Eq.~\eqref{eq.GL} are determined by taking pairing symmetry of both condensates into account.  For detailed derivation of the GL functional we refer to Ref.~\onlinecite{renGinzburgLandauEquationsVortex1995}. In Eq.~\eqref{eq.GL}, the dimensionless units for the order parameters and all distances are $\Delta_0=\sqrt{(4/3\alpha) \ln (T_d/T)}$ and $\xi=v_F\sqrt{\alpha/ \ln(T_d/T)}/2$, respectively, where $\alpha=7\zeta(3)/(8\pi^2 T^2)$. The dimensionless unit for vector potential $\vec{A}$ is $\Phi_0/(2\pi\xi)$ with $\Phi_0=hc/2e$ being the flux quantum.  

Minimizing the free energy, we arrive at the following GL equations,
\begin{align}
\begin{split}
  -\alpha_s \psi_s &+\frac{4}{3}|\psi_s|^2\psi_s +\frac{2}{3}|\psi_d|^2\psi_s+\frac{2}{3}\psi_d^2\psi^*_s \\
  &+ {\Pi^*}^2\psi_s+\frac{1}{2}({\Pi_x^*}^2-{\Pi_y^*}^2)\psi_d = 0,
\end{split}
\\[1ex]
\begin{split}
  -\psi_d &+|\psi_d|^2\psi_d +\frac{8}{3}|\psi_s|^2\psi_d+\frac{4}{3}\psi_s^2\psi^*_d \\
  &+ {\Pi^*}^2\psi_d+({\Pi_x^*}^2-{\Pi_y^*}^2)\psi_s = 0,
\end{split}
\\[1ex]
\begin{split}
  \kappa^2(\nabla \times \nabla \times \vec{A}) &= (\psi_s^*\Pi^*\psi_s+c.c.) \\
  &+\frac{1}{2}(\psi_d^*\Pi^*\psi_d+c.c.) \\
  &+ \frac{1}{2}(\psi_d^*\Pi^*_x\psi_s+\psi_s^*\Pi^*_x\psi_d+c.c.)\hat{x} \\
  &+  \frac{1}{2}(\psi_d^*\Pi^*_y\psi_s+\psi_s^*\Pi^*_y\psi_d+c.c.)\hat{y}.
\end{split}
\end{align}
These equations are then solved numerically and self-consistently using gradient descent algorithm, as a two-dimensional problem with periodic boundary conditions and with applied perpendicular magnetic field. The boundary conditions impose the constraint of fixing the average magnetic field $\bar{H}$ by specifying the integer flux $m=\Phi/\Phi_0$ in the unit cell. The general expression for periodic conditions on a unit cell spanned by vectors $\vec{t_1}$ and $\vec{t_2}$ is\cite{duModelingAnalysisPeriodic1993}.
\begin{align*}
&\psi_{s,d}(\vec{x}+\vec{t}_k)=\psi_{s,d}(\vec{x}) \exp (ig_k(x)),\\
&\vec{A}(\vec{x}+\vec{t}_k)=\vec{A}(\vec{x})+\nabla(g_k(x)),\\
&g_{k}(\vec{x})\equiv c_k-\frac{1}{2}\bar{H}[(1+\theta)t_{ky}x-(1-\theta)t_{kx}y].
\end{align*}
Here, $g_k(x)$ is the so-called generation function, and $\theta$ and $c_k$($k=1,2$) are arbitrary constants. 
The applied magnetic field is given by $\bar{H} = 2\pi m/\Omega$ with integer $m$ being the number of the magnetic flux quanta through the simulated unit cell with area $\Omega = |\vec{t_1}\times\vec{t_2}|$.

Unless specified otherwise, in this paper we consider a large unit cell with area $\Omega = 1024 \xi^2$ and calculate on a dense numerical mesh with grid spacing of $0.1\xi$.  For convenience, we define the unit of magnetic field as $H_0 \equiv  2\pi /\Omega$, so that the applied field can be expressed as $\bar{H} = m H_0$. We also take $\theta=-1$ and $c_1=c_2=0$ for simplicity.  The chosen GL parameter $\kappa=4$ is representative for a type-II superconductor.  We confirmed that all reported features remain robust for $\kappa=2$ and $\kappa \rightarrow \infty$.

The simulations start from multiple artificially generated initial conditions for each value of the parameter $\alpha_s$.  These initial conditions include artificially generated regular vortex configurations and randomly generated vortex states.  Then, we sweep the aspect ratio ($\gamma=|\vec{t_1}/\vec{t_2}|=1$ to $3$) and the magnetic field $\bar{H}=H_0$ to $16H_0$ back and forth until no other new solution is found. After repeating the above procedure many times, the lowest energy state can be regarded as the ground state for each value of $\alpha_s$ and magnetic field $\bar{H}$.

\section{Results}\label{sec:3}

\begin{table}
\begin{center}
 \begin{tabular}{m{2.4cm} m{2cm} m{1.6cm} m{2.2cm}}
 \hline\hline
      & $|\psi_s|^2$ & $|\psi_d|^2$ & $\theta_{sd}$ \\ [0.5ex] 
      \hline
 $\alpha_s \leqslant 2/3 $ & $0$ & $1$ & - \\ [0.5ex] 
 $2/3 < \alpha_s < 1$ & $(9\alpha_s-6)/4$ & $3(1-\alpha_s)$ & $\pm \pi/2$ (TRSB) \\[0.5ex] 
 \hline\hline
\end{tabular}
\end{center}
\caption{\label{tb1} Ground state of the mixed $s$- and $d$-wave GL model in absence of a magnetic field, as a function of the relative strength of the condensates $\alpha_s$.  $|\psi_s|^2$ and $|\psi_d|^2$ are the Cooper-pair densities of the $s$- and $d$-wave condensates, respectively.  $\theta_{sd} =\theta_s-\theta_d$ is the phase difference between the condensates.  The time reversal symmetry breaking (TRSB) $s+id$ state sets in for $\alpha_s \in (2/3, 1)$.}
\end{table}

In absence of applied magnetic field, i.e. for $\nabla\times \vec{A}=0$ and $\Pi\psi_s=\Pi\psi_d=0$, the GL free energy \eqref{eq.GL} is expressed as
\begin{align}\label{eq.GLgs}
F=& -2\alpha_s|\psi_s|^2-|\psi_d|^2+\frac{4}{3}|\psi_s|^4+\frac{1}{2}|\psi_d|^4 \notag\\
&+\underbrace{\frac{8}{3}|\psi_s|^2|\psi_d|^2}_{\eta} +\underbrace{\frac{4}{3}|\psi_s|^2|\psi_d|^2 \mathrm{cos}(2\theta_{sd})}_{\delta}.
\end{align}
The ground-state homogeneous solutions, obtained by minimizing this free energy, are given in Table.~\ref{tb1}.

When $\alpha_s \leqslant \alpha'_s=2/3$, the ground state is a purely $d$-wave state.  In contrast, when $2/3 < \alpha_s < 1$, both condensates have non-zero density.  Due to the Josephson coupling $\delta$ term in Eq.~\eqref{eq.GLgs}, the condensates favor a phase difference $\theta_{sd}=\pm\pi/2$, forming the $s\pm id$ state.  In that phase, there are two degenerate uniform phases, i.e. the $s+id$ state and the $s-id$ state.  One can never transform either of them into the other by the U(1) gauge transformation, thus leading to the TRSB.  However, the system remains invariant under combination of time-reversal symmetry operation and $C_4$ rotations.

\begin{figure}[t]
	\centering
	\includegraphics[width=\columnwidth]{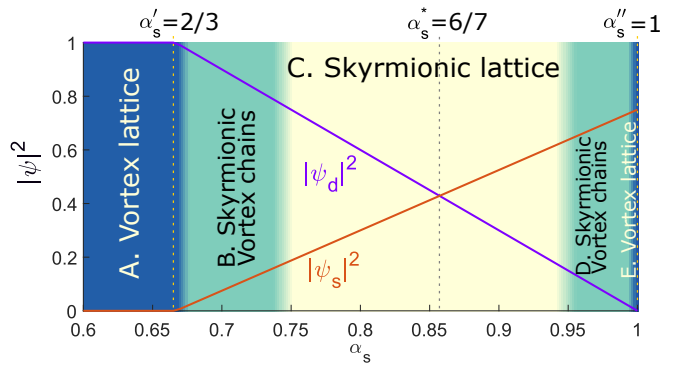}
	\caption{(Color online) Solid lines show the homogeneous Cooper-pair density of the $s-$ and $d-$ condensates, $|\psi_s|^2$ and $|\psi_d|^2$, as a function of their relative strength $\alpha_s$, in absence of magnetic field.  $\alpha'_s=2/3$ and $\alpha''_s=1$ are the critical values for which $|\psi_s|$ and $|\psi_d|$ vanish, respectively.  At $\alpha^*_s=6/7$, $|\psi_d|=|\psi_s|$.  Depending on the relative amplitude $|\psi_d/\psi_s|$, i.e. $\alpha_s$, the $s+id$ state under magnetic field can exhibit vortex lattice, skyrmionic chains or skyrmionic lattices (depicted by different colors and transitions between them).}
	\label{fig.gs}
\end{figure}

Note that in the $s + id$ state regime, the density of both condensates, $|\psi_d|$ and $|\psi_s|$, changes with variation of their relative strength $\alpha_s$, as shown in Fig.~\ref{fig.gs}.  One finds a transition from a purely $d$-wave state at $\alpha'_s = 2/3$ to a purely $s$-wave state at $\alpha''_s = 1$.  At $\alpha^*_s = 6/7$, both condensates have the same density, i.e. $|\psi_s| = |\psi_d|=\sqrt{3/7}$. Therefore, the competition between the two condensates changes with $\alpha_s$.

In what follows, we show that in the $s+id$ state the system under magnetic field undergoes transitions between five vortical states, i.e. states A-E shown in Fig.~\ref{fig.gs}, with increasing $\alpha_s$, or equivalently, $|\psi_d/\psi_s|$.   These states are further classified into three distinct classes, namely the vortex lattice (states A and E), skyrmionic vortex chains (B and D), and skyrmionic lattices (C).  States D and E are analogous to the states B and A, respectively, upon swapping the role between the $s$-wave and the $d$-wave condensates.  Therefore, in the remainder of the article we focus on the states A-C, being representative of the three classes of characteristic vortical states in a $s+id$ superconductor.

\subsection{Vortex lattice at the onset of the $s+id$ state}\label{sec:3.1}

\begin{figure*}
	\centering
	\includegraphics[width=0.8\textwidth]{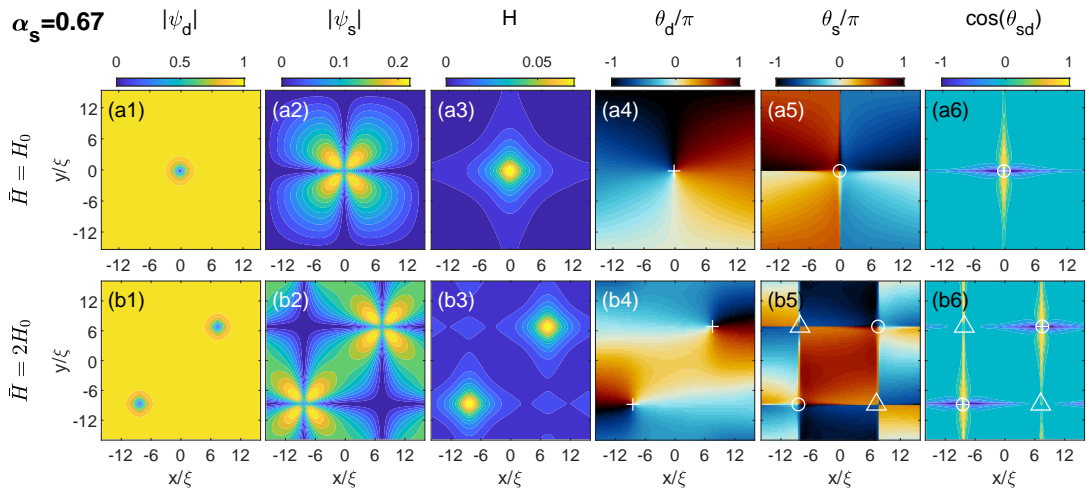}
	\caption{(Color online) Structure of a single vortex (upper panels) and the square vortex lattice (lower panel) for $\alpha_s =0.67$, obtained by threading one and two flux quanta through the simulation unit cell, respectively.  Panels left-to-right show the spatial distributions of the $d$-wave amplitude $|\psi_d|$, $s$-wave amplitude $|\psi_s|$, magnetic field intensity $H$, the phase distributions of the $d$-wave $\theta_d$, $s$-wave $\theta_s$, and the phase difference $\theta_{sd}$ between the two condensates, respectively.  The cross sign indicates a vortex with phase winding $+1$ in the $d$-wave component $\psi_d$.  Open circles and open triangles indicate vortices with phase winding $-1$ and $+2$, respectively, in the $s$-wave component $\psi_s$.}
	\label{fig.vortex1}
\end{figure*}

Vortex states in the $s+d$-wave regime ($\alpha_s \leqslant \alpha'_s$) have been well studied previously.  In the ground state, the density is zero in the $s$-wave component $|\psi_s|$.  However, near a vortex core in the $d$-wave component $\psi_d$, the $s$-wave component is always induced, and has an opposite winding number relative to the $d$-wave component.  As $\alpha_s$ approaches $\alpha'_s$ from below, the induced $s$-wave component near the vortex core becomes more significant.  This results in the change in the vortex structure from isotropic to the fourfold anisotropic, affecting further the vortex lattice and inducing the change from triangular to square lattice \cite{xuStructuresSingleVortex1996}.

When $\alpha_s > \alpha'_s$, the system enters the $s+id$-wave regime.  Both $s$-wave and $d$-wave components are nonzero in the ground state.  However, when $\alpha_s$ is very close to the critical $\alpha'_s$, the density of the $s$-wave component is low.  Therefore, the vortex states remain the same as in the $s+d$-wave regime.  In this section, we show vortex states for $\alpha_s=0.67$ ($\approx \alpha'_s$).  It results in the $d$-wave component density $|\psi_d| \rightarrow 1$ whereas the $s$-wave component density $|\psi_s| \approx 0.1$.  We shall review some important features of the vortex structure for facilitated understanding of the vortex (skyrmionic) chains in the next section. 

Fig.~\ref{fig.vortex1}(a1)-(a6) focuses on the structure of a single vortex.  The vortex is located in the center of the unit cell, where the magnetic field intensity $H$ reaches its maximum [Fig.~\ref{fig.vortex1}(a3)]. As seen, the $d$-wave component $\psi_d$ contains a vortex with phase winding $+1$ [Fig.~\ref{fig.vortex1}(a4)], and density $|\psi_d|$ drops to zero in the vortex core.  For such a $d$-wave component $\psi_d$, the $s$-wave component $\psi_s$ has the solution of the form $\psi_s \sim f_1(r)e^{-i\theta} + f_2(r)e^{i3\theta}$, when $|\psi_d| \gg |\psi_s|$ and $|\nabla \psi_d| \gg |\nabla \psi_s|$.\cite{franzVortexStateDwave1996} It means that there is an antivortex inside the core, whose winding number is $-1$.  This is induced by the $x-y$ asymmetric form of the mixed gradient term [the $\zeta$ term in Eq.~\eqref{eq.GL}].  Therefore, the $s$-wave component $\psi_s$ contains an antivortex, coinciding with the positive vortex in the $d$-wave component $\psi_d$.

The opposite winding between two components leads to their relative phase $\theta_{sd}$ twirling twice around the vortex core, with four domain walls appearing along the $\pm x$ and $\pm y$ axes [Fig.~\ref{fig.vortex1}(a6)], on which either $\theta_{sd}=0$ or $\pi$.   The domain walls are energetically unfavorable because $\theta_{sd}$ on the domain walls deviate from the preferred value $\pm \pi/2$ stemming from the Josephson coupling [$\delta$ term in Eq.~\eqref{eq.GL}].  In regions outside the domain walls, the relative phase $\theta_{sd}$ is indeed $\pm \pi/2$.

To lower the energy cost of the domain wall, the phase change across the domain walls is very rapid, such that the density in the $s$-wave component $|\psi_s|$ is suppressed there.  In this way, the $s$-wave component $|\psi_s|$ exhibits a four-leafed clover shape [Fig.~\ref{fig.vortex1}(a2)], which further leads to the fourfold symmetry reflected in the $d$-wave component $|\psi_d|$ and in the magnetic field intensity $H$ far away from the vortex core.  Note that the phase change on each domain wall is $\pi$.  The four domain walls contribute total phase winding $+2$ at the boundary of the unit cell.  As a result, the total phase winding in the $s$-wave component $\psi_s$ is $+1$, the same as in the $d$-wave component $\psi_d$.

\begin{figure*}
	\centering
	\includegraphics[width=0.8\textwidth]{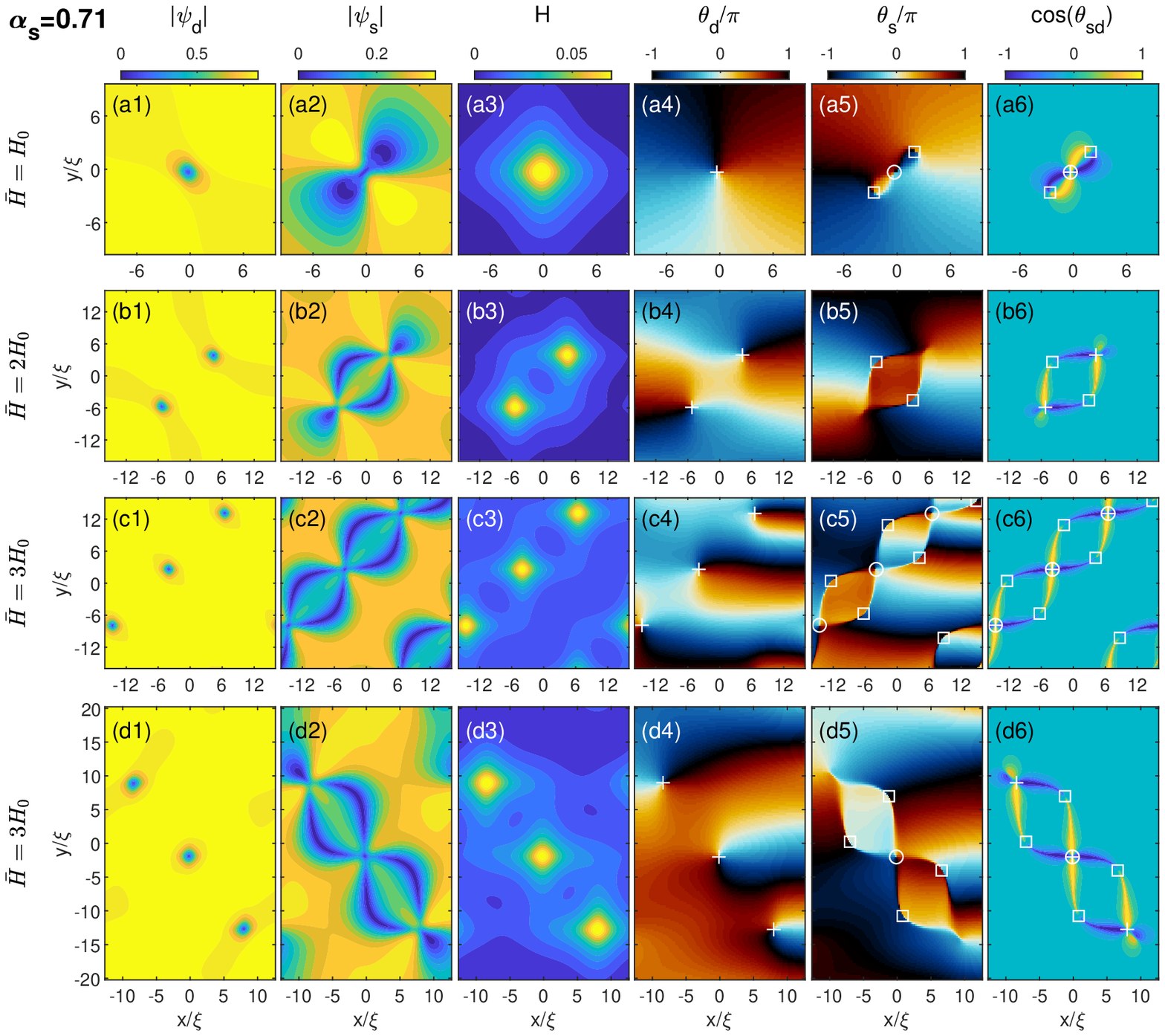}
	\caption{(Color online) Structure of a single vortex (top row), a dimer vortex state (the second row), a continuous vortex chain (the third row) and a terminated vortex chain (the last row), obtained by threading one, two, three and three flux quanta, respectively, through the unit cell.  The first three cases are obtained using a square unit cell, while the last case is obtained using a rectangular unit cell with aspect ratio $\gamma=1.8$. Panels from left to right show the spatial distributions of the $d$-wave amplitude $|\psi_d|$, $s$-wave amplitude $|\psi_s|$, magnetic field intensity $H$, the phase distributions of the $d$-wave $\theta_d$, $s$-wave $\theta_s$, and the phase difference $\theta_{s,d}$ between the two condensates, respectively.  The $+$ sign indicates a vortex with phase winding $+1$ in $d$-wave component $\psi_d$.  Open circle and open square indicate a vortex with phase winding $-1$ and $+1$, respectively, in the $s$-wave component $\psi_s$.}
	\label{fig.chain1}
\end{figure*}

To illustrate the features of the vortex lattice rather than the isolated vortices, we threaded two flux quanta ($m=2$) through the unit cell.  By comparing the free energy for various aspect ratios $\gamma$ of the unit cell, we found that the minimum free energy is obtained when $\gamma=1$, i.e. for the square lattice. The structure of that lattice is shown in Fig.~\ref{fig.vortex1}(b1)-(b6).  As seen from the $d$-wave component $|\psi_d|$ [Fig.~\ref{fig.vortex1}(b1)] and the magnetic field $H$ [Fig.~\ref{fig.vortex1}(b3)], two vortices are located on the diagonal of the unit cell. The $s$-wave component $|\psi_s|$ and the relative phase $\theta_{sd}$ are arranged in the same way.  

One should note that the total $+2$ state winding in the $s$-wave component $\psi_s$ [see Fig.~\ref{fig.vortex1}(b5)] is reached via a square sublattice of vortices with $+2$ state winding, with two antivortices residing at locations of vortices in the $d$-wave component $\psi_d$. However, the giant-vortex-like $+2$ state winding is energetically expensive. Therefore, when the density in the $s$-wave $|\psi_s|$ is further enhanced with $\alpha_s$ increasing, the vortices with phase winding $+2$ will become unstable, leading to completely different vortex states.

\subsection{Skyrmionic vortex chains}\label{sec:3.2}

When $\alpha_s$ is increased, the $s$-wave component $\psi_s$ becomes more important, and couples back to the $d$-component more strongly.  To start with, it changes the structure of a single vortex, as compared to the case in the previous section.  In addition, the vortices favor to be interconnected via phase domain walls, forming vortex chains that exhibit skyrmionic character.  In this section, we present results for $\alpha_s =0.71$ as an example.  In the ground state, the density of the $d$-wave component is $|\psi_d| =0.93$ while the density in the $s$-wave component is $|\psi_s| = 0.31$, making the relative amplitude $|\psi_d/\psi_s| = 3$.

Fig.~\ref{fig.chain1}(a1)-(a6) show the single vortex state, obtained by threading one flux quantum in the square unit cell.  The single vortex is at center, i.e. $(x,y)=(0,0)$, where the dominant $d$-wave component contains a vortex with phase winding $+1$, and where the peak in the magnetic field intensity $H$ is found [Fig.~\ref{fig.chain1}(a3)].  As before, the $s$-wave component $\psi_s$ carries the opposite winding relative to the $d$-wave component $\psi_d$ near the vortex core due to the mixed gradient term [the $\zeta$ term in Eq.~\eqref{eq.GL}].  It means that the antivortex with phase winding $-1$ in the $s$-wave component $\psi_s$ is superimposed on a positive vortex in $\psi_d$. Therefore, this is a composite vortex where both components' densities drop to zero in the vortex core.  

\begin{figure*}
	\centering
	\includegraphics[width=0.95\textwidth]{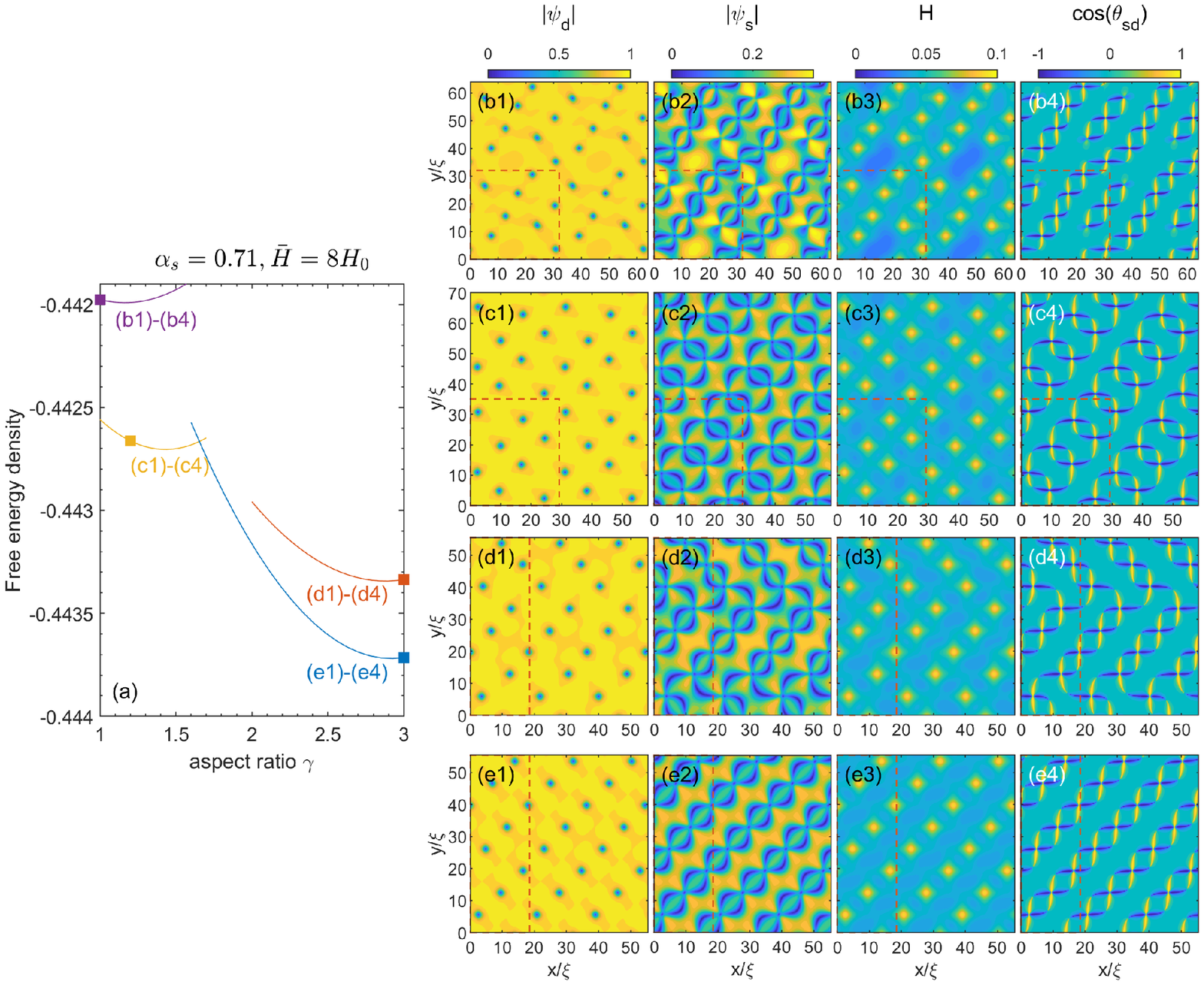}
	\caption{(Color online) Panel (a) shows the free energy of four typical vortex states found for different aspect ratio of the simulation region (dashed box), for $\alpha_s =0.71$ and 8 times larger magnetic field than in Fig.~\ref{fig.chain1}(a).  The vortex configurations marked by solid squares are shown in panels (b), (c), (d) and (e), respectively.  Panels labeled 1-4 show the $d$-wave amplitude $|\psi_d|$, $s$-wave amplitude $|\psi_s|$, magnetic field intensity $H$, and the phase difference $\theta_{sd}$, respectively.}
	\label{fig.chain2}
\end{figure*}

Different from the previous case where there were four positive vortices around the antivortex in the $s$-wave component $\psi_s$, there are only two positive vortices around the antivortex in this case, as seen in the phase plot in Fig.~\ref{fig.chain1}(a5), and the trace in the density of the $s$-wave component in Fig.~\ref{fig.chain1}(a2). The two positive vortices in the $s$-wave component $\psi_s$ are connected with the composite vortex at $(0,0)$ through phase domain walls, with a phase knot at the composite vortex  [Fig.~\ref{fig.chain1}(a6)].  In fact, the domain walls become significantly shorter as compared to the previous case.  This is because the domain walls cost more energy in this case as $|\psi_s|^2|\psi_d|^2$ becomes larger in the Josephson coupling $\delta$ term in Eq.~\eqref{eq.GL}.  As a result, the domain walls shrink for saving energy and pull the attached positive vortex in the $s$-wave component $\psi_s$ from the boundary of the unit cell to a position closer to the composite vortex. 

\begin{figure*}
	\centering
	\includegraphics[width=0.8\textwidth]{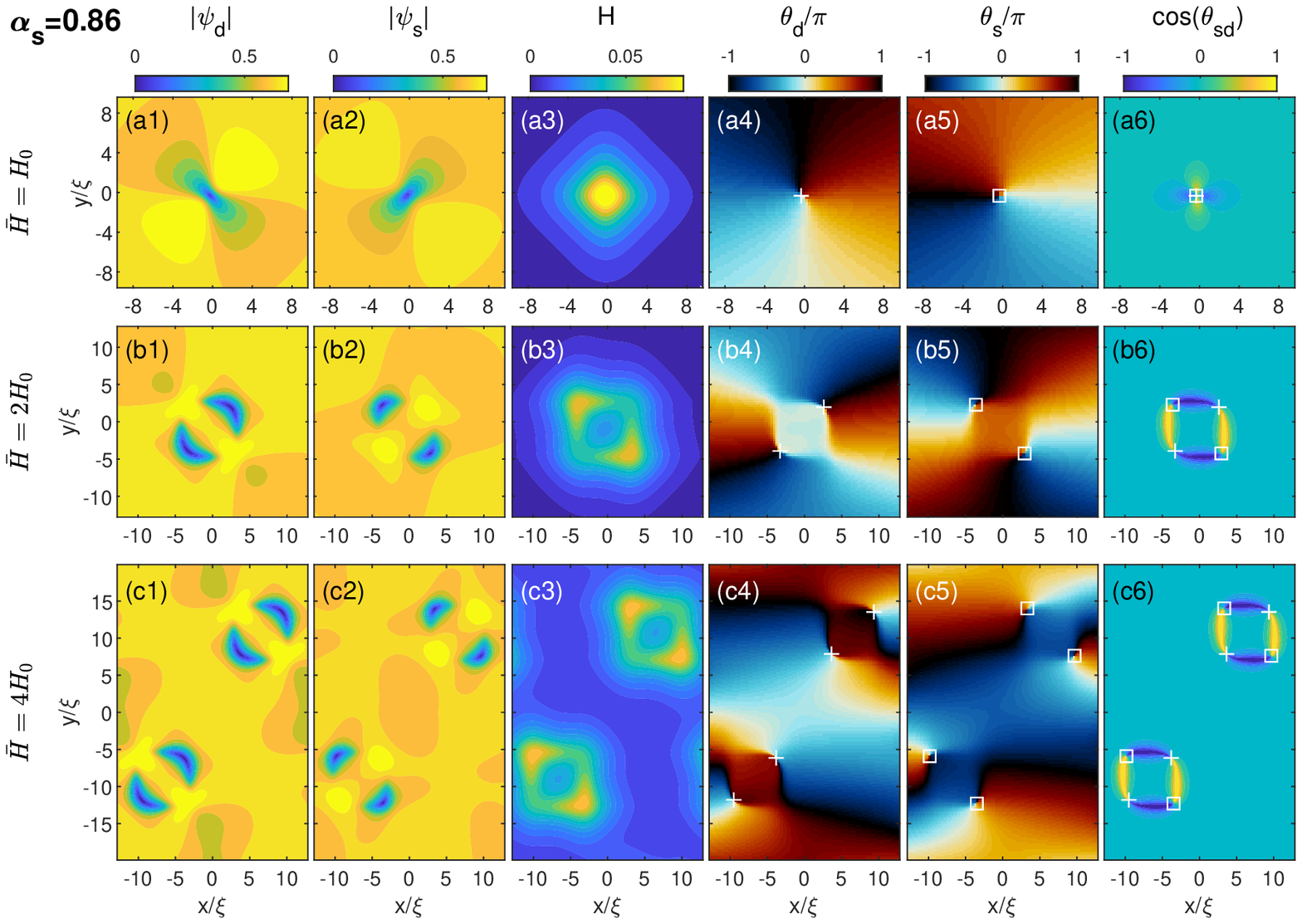}
	\caption{(Color online) Structure of a single vortex (a), doubly quantized coreless vortex, i.e. skyrmion (b) and doubly quantized coreless vortices in a lattice (c), for $\alpha_s =0.86$, obtained by threading one and two flux quanta through the square unit cell, and four flux quanta in the rectangular unit cell with aspect ratio $\gamma=1.5$, respectively.  Panels from left to right show the spatial distributions of the $d$-wave amplitude $|\psi_d|$, $s$-wave amplitude $|\psi_s|$, magnetic field intensity $H$, the phase distributions of the $d$-wave $\theta_d$, $s$-wave $\theta_s$, and the phase difference $\theta_{sd}$ between the two condensates, respectively. $+$ marks and open squares indicate vortices with phase winding $+1$ in $\psi_d$ and $\psi_s$, respectively.}
	\label{fig.skyr1}
\end{figure*}

To understand the interaction of such vortices, we next consider the states with two flux quanta in the unit cell, with structure shown in Fig.~\ref{fig.chain1}(b1)-(b6).  As seen, the two positive vortices in the $d$-wave component $\psi_d$ are on a diagonal of the unit cell [indicated by $+$ marks in Fig.~\ref{fig.chain1}(b4)], accompanied by two maxima in the magnetic field intensity $H$ there [Fig.~\ref{fig.chain1}(b3)].  Interestingly, the $s$-wave component $\psi_s$ has two positive vortices on the other diagonal of the unit cell [open squares in Fig.~\ref{fig.chain1}(b5)], that connect to the vortices in $\psi_d$ via phase domain walls [Fig.~\ref{fig.chain1}(b6)]. The density in the $s$-wave component, $|\psi_s|$, is low on the domain walls, forming the same ring pattern. The corresponding ring-shape is visible also in the magnetic field intensity $H$ at the outskirt of the two main peaks [Fig.~\ref{fig.chain1}(b3)]. Therefore, we see that the structure of an isolated vortex in Fig.~\ref{fig.chain1}(a6) is not preserved when higher density of vortices is imposed. Instead of the separated vortex-antivortex eight-loops, one loop is formed containing all the singularities (with one vortex-antivortex pair in the $s$-component annihilating under the vortices in the $d$-component, but with phase trace remaining due to the mixed gradient term in Eq.~\eqref{eq.GL}), with phase difference $\pi/2$ inside the loop and $-\pi/2$ outside. As will be discussed further on, the corresponding domain wall exhibits skyrmionic topology, similarly to the cases of split-vortices in multiband and multicomponent superconductors in the literature.\cite{zhang_electronic_2016, garaudPropertiesSkyrmionsMultiquanta2015} 

With further increasing vortex density, the above discussed loops can interconnect throughout the sample, and form skyrmionic vortex chains.  Here the antivortex in $\psi_s$ again coincides with the positive vortex in the $\psi_d$ to form the composite vortex, and the additional vortices in $\psi_s$ attach to domain walls to connect the chain. Inside the chain the phase difference between the components is $\pi/2$, and $-\pi/2$ outside the chain. Fig.~\ref{fig.chain1}(c1)-(c6) shows the structure of such a skyrmionic vortex chain.  As seen, the three composite vortices are connected through domain walls, forming a chain.  They are arranged to form a line, that results in three aligned peaks in the magnetic field intensity $H$, with visible outskirt structure due to domain walls. 

A skyrmionic vortex chain can also terminate and take finite (short) length [see Fig.~\ref{fig.chain1}(d)], by annihilating vortex-antivortex pairs in $\psi_s$ at its ends.  It retains similar features to what was seen in the continuous (long) vortex chain [see Fig.~\ref{fig.chain1}(b)], but also combines features from the dimer state [see Fig.~\ref{fig.chain1}(c)].  However, the short, terminated vortex chains are always energetically more expensive than the continuous (long) vortex chains.  

To find the ground state at different magnetic fields, we have performed simulations by increasing the number of flux quanta piercing the unit cell up to $m=16$.  We find that the continuous vortex chains [shown in Fig.~\ref{fig.chain1}(c1)-(c6)] always have the lowest energy. Therefore, the skyrmionic vortex chains are energetically favorable and observable. To prove the point, we show in Fig.~\ref{fig.chain2} four typical vortex states obtained for the magnetic field 8 times larger than in Fig.~\ref{fig.chain1}(a), where we varied the aspect ratio of the simulation region in order to probe all possible stable configurations.  The free energies of found states are shown in panel (a), proving that the lowest-energy state is indeed the one with vortex chains [Fig.~\ref{fig.chain1}(e1)-(e4)], followed by zig-zag vortex chains [Fig.~\ref{fig.chain1}(d1)-(d4)], double vortex chains [Fig.~\ref{fig.chain1}(c1)-(c4)], and the fractional chain state [Fig.~\ref{fig.chain1}(b1)-(b4)]. 

Chain organization of vortices indicates presence of the non-monotonic interactions and strong multi-body forces between them. In a conceptually simpler two-component system where two condensates are only coupled by density coupling (the same as the $\eta$ term in Eq.~\eqref{eq.GL} but with a different prefactor), the tendency to form vortex chains is found in the phase-separated ground state (i.e. $|\psi_1| \neq 0$ but $|\psi_2| =0$).\cite{garaudVortexChainsDue2015}  However, it competes with the increased importance of current-current interactions in the relatively dense vortex matter.  Therefore, the vortex chains in that case will become less ordered under higher magnetic field. In our $s+id$ state, in contrast, the vortex chain is very stable even under high magnetic field where the vortex matter is dense. This is attributed to the opposite phase winding between two condensates near the composite vortex, in which the domain walls are knotted. The opposite winding between components inside the composite vortex also generates additional single-component vortices, that support propagation of domain walls through the sample. In such a way, composite vortices interconnect to form skyrmionic chains and minimize the energy.

\subsection{Skyrmionic lattices}\label{sec:3.3}

In what follows, we examine the further trend in the vortical configurations as $\alpha_s$ is increased further and the $s$-wave component $|\psi_s|$ in the ground state becomes comparable to the $d$-wave component. In this regime, we find that each vortex tends to split its core within different condensates, and form coreless vortex structures that are characterized by a skyrmionic topological structure.  In this section, we exemplify the results for $\alpha_s=0.86$, where the ground state amplitudes of the components are $|\psi_s| \approx 0.66$ and $|\psi_d| \approx 0.65$.

Fig.~\ref{fig.skyr1}(a1)-(a6) shows the structure of an isolated vortex.  The $s$-wave component $\psi_s$ exhibits a large change as compared to the case shown in Fig.~\ref{fig.chain1}(a1)-(a6) as the vortex-antivortex molecule (best seen in Fig.~\ref{fig.chain1}(a5)) recombines into just one positive vortex, forming a composite vortex with the vortex of same winding in the $d$-wave component.  Such a composite vortex is a consequence of the Josephson coupling $\delta$ term and the direct coupling $\eta$ term in Eq.~\eqref{eq.GL} overcoming the gradient $\zeta$ term, so that opposite local phase winding in two condensates is no longer warranted.

The obtained single-vortex structure is similar to one found in chiral $p$-wave superconductors.\cite{zhang_electronic_2016}  Both components, $|\psi_s|$ and $|\psi_d|$, have nearly same spatial distribution and winding symmetry.  In other words, the $s$-wave component $\psi_s$ can be nearly obtained by rotating the $d$-wave component $\psi_d$ by $90^{\circ}$.  As a result, the phase difference $\theta_{sd}$ is near the energetically favorable value $\pm \pi/2$ in the entire region, even close to the vortex core.

Similar to the chiral $p$-wave superconductor, the $s+id$ state in this case favors to form coreless vortices, i.e. clusters of vortices with split cores between the component condensates. This is seen upon increasing the magnetic field, as shown for $m=2$ (two flux quanta in the unit cell) in Fig.~\ref{fig.skyr1}(b1)-(b6). Instead of the vortex state with two composite vortices, we find that it is energetically cheaper to have vortex cores spatially separated, as facilitated by the direct coupling term $\eta$. Namely, the two vortices in the $d$-wave component $\psi_d$ [see symbols $+$ in Fig.~\ref{fig.skyr1}(b)] are located on the diagonal of the unit cell while the two vortices in the $s$-wave component $\psi_s$ (open squares in Fig.~\ref{fig.skyr1}(b)) are on the other diagonal.  The four one-component vortex cores therefore form a loop structure, interconnected by domain walls in the relative phase [Fig.~\ref{fig.skyr1}(b6)].  Such structure hosts $\theta_{sd}$ being $\pi/2$ inside, and $-\pi/2$ outside the domain wall.  As seen previously, the domain wall traps vortices to lower its energy cost.  The magnetic field intensity $H$ exhibits a corresponding ring structure [Fig.~\ref{fig.skyr1}(b3)], with peaks found at the vortex cores of the $s$-wave component $\psi_s$ (being slightly dominant for $\alpha_s=0.86$).

\begin{figure}
	\centering
	\includegraphics[width=\columnwidth]{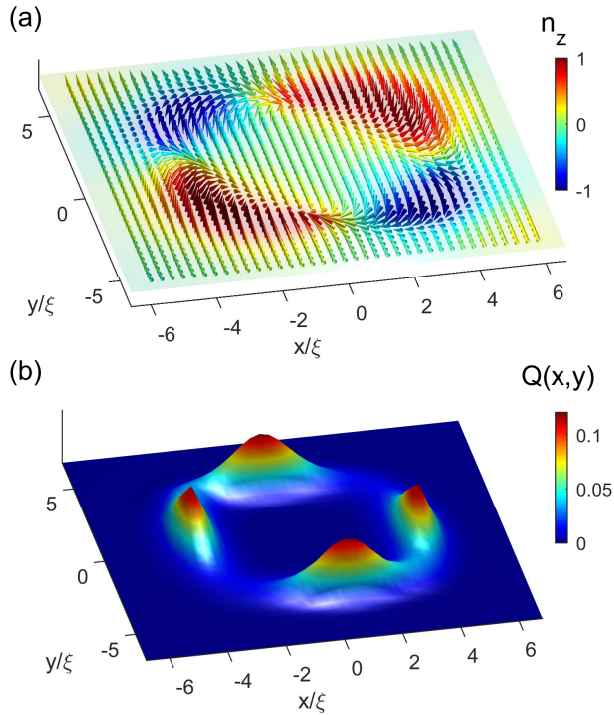}
	\caption{(Color online) (a) The pseudo-spin texture $n(x,y)$ of the double-quanta coreless vortex shown in Fig.~\ref{fig.skyr1}(b1)-(b6).  The color indicates the $z$ amplitude of the pseudo-spin field $\mathbf{n}(x,y)$. (b) Topological charge density $Q(x,y)$.  It is non-trivial on the closed domain wall.  This double-quanta coreless vortex exhibits a topological charge $\mathbb{Q}=2$, and is therefore a skyrmion.}
	\label{fig.texture}
\end{figure}

\begin{figure*}
	\centering
	\includegraphics[width=0.95\textwidth]{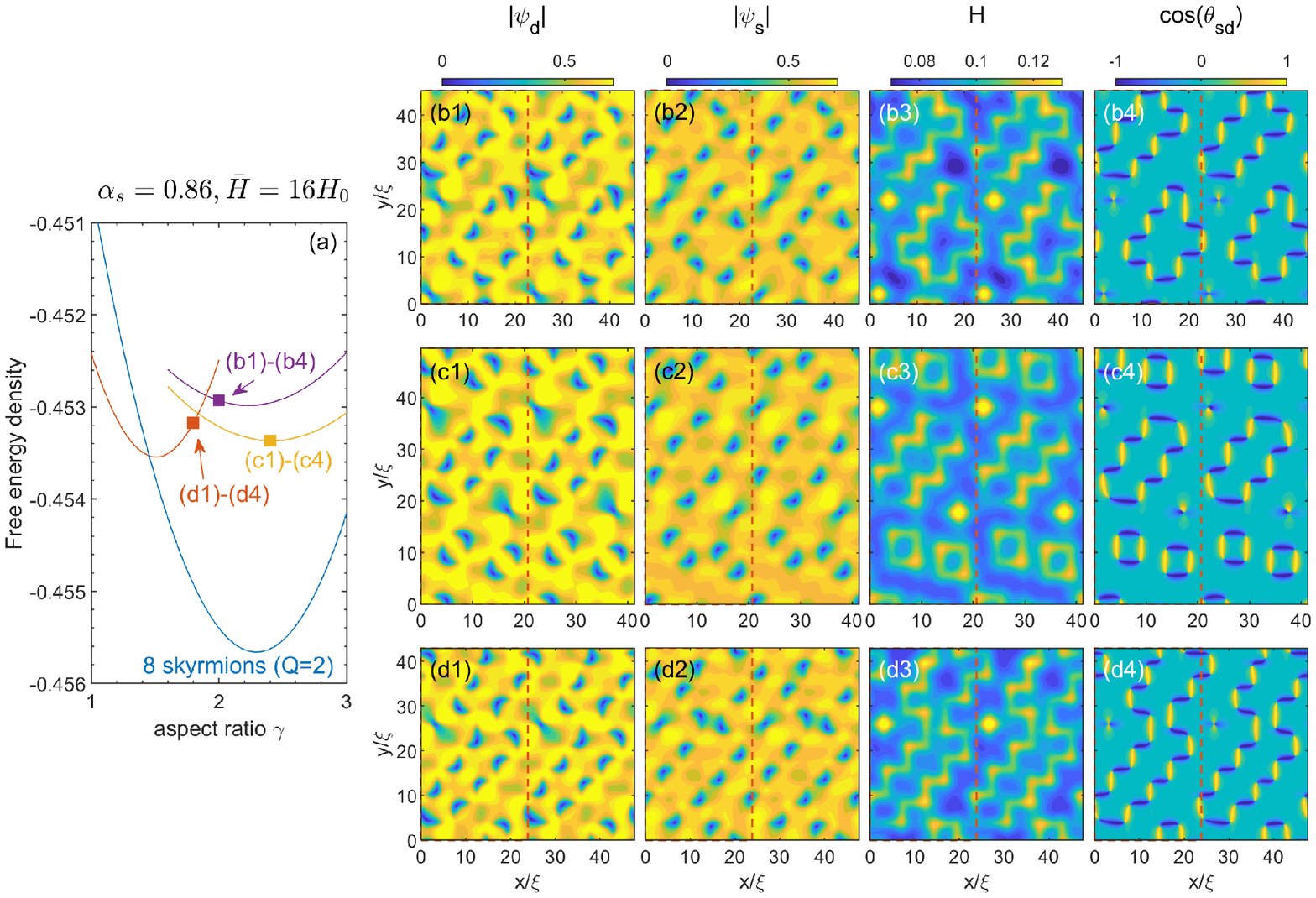}
	\caption{(Color online) Panel (a) shows the free energy of four typical vortex states as a function of aspect ratio $\gamma$ for $\alpha_s =0.86$ and $16$ flux quanta threading the simulation region (dashed box).  The structures indicated by solid squares in (a) are shown in panels (b1)-(b4), (c1)-(c4) and (d1)-(d4), respectively.  The state shown in panels (b1)-(b4) contains $\mathbb{Q}=7$ skyrmions and single-quanta vortices.  The state shown in panels (c1)-(c4) contains $\mathbb{Q}=6$ skyrmions, $\mathbb{Q}=2$ skyrmions, and single-quanta vortices.  The state shown in panels (d1)-(d4) contains $\mathbb{Q}=15$ skyrmions and single-quanta vortices.  Panels labeled 1-4 show the $d$-wave amplitude $|\psi_d|$, $s$-wave amplitude $|\psi_s|$, magnetic field intensity $H$ and the phase difference $\theta_{sd}$, respectively.  The lowest energy state is the one with a pure lattice of $\mathbb{Q}=2$ skyrmions (configuration not shown).}
	\label{fig.skyr16}
\end{figure*}

This structure of four one-component vortices alternating on a phase domain wall can be seen as a double-quanta coreless vortex, but is topologically a representation of a skyrmion. To reveal the skyrmionic feature of the double-quanta coreless vortex, we map the two-component order parameter $\psi_s$ and $\psi_d$ on the pseudo-spin $\mathbf{n}$.  The $\mathbf{n}$ is defined as\cite{babaevHiddenSymmetryKnot2002}: 
\begin{equation}\label{Eq:pesudospin}
\mathbf{n} = (n_x, n_y, n_z) = \frac{\mathbf{\psi}^\dagger \vec{\sigma} \mathbf{\psi}}{\mathbf{\psi}^\dagger \mathbf{\psi}},
\end{equation}
where $\mathbf{\psi} = (\psi_s, \psi_d)$ and $\vec{\sigma}$ is the Pauli matrices.  $\mathbf{n}$ is a 3D unit vector $|\mathbf{n}|=1$, and points to the surface of a solid unit sphere in three-dimensional space $S^2$.  The pseudo-spin texture is shown in Fig.~\ref{fig.texture}(a).  Notably, the $n_{z}$ projection of the pseudo-spin flips four times when moving along the domain wall.  This indicates that $\mathbf{n}$ wraps twice inside the unit cell.  To clarify this further, we also calculate the topological charge $\mathbb{Q}$ as
\begin{equation}\label{Eq:TopoCharge}
    \mathbb{Q}=\int Q(x,y) dxdy = \frac{1}{4\pi} \int \mathbf{n} \cdot (\partial_x \mathbf{n} \times \partial_y \mathbf{n}) ~ dxdy,
\end{equation}
where $Q(x,y)$ is the topological charge density.  As shown in Fig.~\ref{fig.texture}(b), the topological charge density $Q(x,y)$ is non-trivial along the domain wall, and the integration on the topological charge density $Q(x,y)$ in space leads to $\mathbb{Q}=2$.  Therefore, the double-quanta coreless vortex is indeed a skyrmion, with topological charge $\mathbb{Q}=2$.

As seen from Fig.~\ref{fig.texture}(b), the topological charge density $Q(x,y)$ is not significant at the four one-component vortex cores.  Instead, it is largest along the domain wall between the vortices. We emphasize again that these skyrmions have a twofold symmetry due to present disbalance between the two components.  They are therefore different from the skyrmions in the chiral $p$-wave superconductor, which possess a cylindrical symmetry.\cite{zhang_electronic_2016}

In the next step, we consider the behavior of above $\mathbb{Q}=2$ skyrmions as magnetic field is increased. We find that lattice of such skyrmions remains favorable compared to states containing single-quanta vortices.  For example, in the case shown in Fig.~\ref{fig.skyr1}(c1)-(c6) for 4 flux quanta in the simulation region, the obtained lowest energy state is the one with a triangular lattice of $\mathbb{Q}=2$ skyrmions! We have tested this further at 4 times larger field, with the same conclusion.  The free energy of the skyrmion lattice state is shown in Fig.~\ref{fig.skyr16}(a) as the lowest-energy state but its configuration is not shown explicitly.

In fact, the skyrmion can have a larger topological charge, i.e. $\mathbb{Q}>2$ by trapping more vortices on the phase domain wall.  Those are usually energetically expensive so the associated vortical configurations are meta-stable. In Fig.~\ref{fig.skyr16} we show the structure of three such meta-stable states and their energies.  The state shown in Fig.~\ref{fig.skyr16}(b1)-(b4) consists of two large skyrmions with $\mathbb{Q}=7$ (seven flux quanta) and two single-quanta vortices in the unit cell.  The state shown in Fig.~\ref{fig.skyr16}(d1)-(d4) consists of a very large skyrmion ($\mathbb{Q}=15$, fifteen flux quanta) and a single-quanta vortex within the unit cell. As seen from Fig.~\ref{fig.skyr16}(a), their free energies are significantly higher than the lowest-energy state with lattice of $\mathbb{Q}=2$ skyrmions.  From stability point of view, skyrmions with arbitrary integer $\mathbb{Q}$ are possible, and if formed, they are protected by a finite potential barrier. One can therefore expect that in a real experimental system, characterized by disorder, finite size, shape anisotropy, this skyrmionic phase will energetically tend to form the lattice of $\mathbb{Q}=2$ skyrmions, but likely coexisting with single-quanta vortices and skyrmions of larger topological charge ($\mathbb{Q}>2$).

\section{Distinction to vortex states in $s+is$ superconductors}\label{sec:4}

The general GL model for $s+is$ superconductors is written as\cite{maitiSpontaneousCurrentsSuperconductor2015}
\begin{align}\label{eq.GLsis}
F&= \frac{1}{\Omega} \int \bigg\{ \sum_{i=1,2}\Big( k_i|\Pi\psi_i^{\ast}|^2 + \alpha_i |\psi_i|^2 + \beta_i|\psi_i|^4 \Big) \notag\\
&+\zeta \Big( \Pi{_{x}^{\ast}}\psi_1\Pi_x\psi_2^{\ast}+\Pi_y^{\ast}\psi_1\Pi_y\psi_2^{\ast} + c.c.\Big) \notag\\
& \gamma|\psi_1|^2|\psi_2|^2 +\delta (\psi_1^{\ast2}\psi_2^2+c.c.) +\kappa^2(\nabla\times \vec{A})^2\bigg\} d\Omega.
\end{align}
Note that the bilinear Josephson term $\psi_1^{\ast}\psi_2+c.c.$ is excluded for stabilizing the $s+is$ state with the relative phase $\theta_{12}=\pi/2$. 
This model is relevant to describe some iron-based superconductors, e.g. the hole-doped Ba$_{1-x}$K${_x}$Fe$_2$As$_2$.\cite{garaudMicroscopicallyDerivedMulticomponent2017}  

The form of the energy functional is the same as in Eq.~\eqref{eq.GL} for the $s+id$ state.  The crucial distinction is that the sign between the $x$ and $y$ part of the mixed gradient term $\zeta$ is changed. The mixed gradient term is symmetric ($\zeta_{x}=\zeta_{y}$) for the $s+is$ state, but it is anti-symmetric $\zeta_{x}=-\zeta_{y}$ for the $s+id$ state.  As a result, the $s+is$ state preserves crystal symmetries but only breaks the time-reversal-symmetry.  In contrast, the $s+id$ state only remains invariant under $C_4$ rotations and subsequently the time-reversal operation.

Both $s+is$ and $s+id$ states support rich topological defects such as vortices, domain walls and skyrmions,\cite{garaudMicroscopicallyDerivedMulticomponent2017} hence bare appearance of these topological defects will not distinguish between those pairing symmetries.  However, the different symmetry in the two states results in different structures of these topological defects, further leading to different magnetic field profile.  The latter can be used for distinguishing between two states experimentally.

To begin with, the magnetic field profile of a (composite) vortex in the $s+id$ state is square or rectangular, as seen from Fig.~\ref{fig.vortex1}, Fig.~\ref{fig.chain2} and Fig.~\ref{fig.skyr16}.  In contrast, the profile of the vortex in the $s+is$ state is circular or elliptical, as seen from e.g. Fig.~4 in Ref.~\onlinecite{garaudDomainWallsTheir2014}.  

For some parameter sets, vortices in both states favor to split cores and attach to domain walls to form skyrmionic phases.  In such cases, the magnetic signature of the vortices is smeared out along the domain wall. For the $s+is$ state, the domain walls enclose in geometry with rounded corners, which results in the same pattern in the magnetic field profile.  As a result, the field profile looks similar to bead strings, as seen from from Fig.~4 in Ref.~\onlinecite{garaudDomainWallsTheir2014}.  In contrast, the domain walls in the $s+id$ state exhibit zig-zag patterns [see Fig.~\ref{fig.skyr16}].  When domain walls enclose (form a skyrmion), the magnetic field profile clearly exhibits squared features.

In addition, the composite vortex in the $s+id$ state intrinsically knots domain walls due to the opposite winding between $s-$ and $d-$ condensates.  This enables rich configurations from square vortex lattice to zigzag chain organizations, with skyrmionic character. The vortex chains may terminate or further connect with other chains through domain walls to form irregular patterns.  In contrast, the composite vortex in the $s+is$ state minimizes its energy on a domain wall, but does not support the knot in domain walls because the winding in two $s-$components is the same.  To test this comparison, we solved Eq.~\eqref{eq.GLsis} for the $s+is$ state with the coefficients taken same as in Eq.~\eqref{eq.GL}, and obtained a conventional triangular vortex lattice, in which each single vortex has a circular symmetry.

\section{Conclusions}\label{sec:5}

In summary, motivated by the lack of distinctive properties of $s+is$ and $s+id$ states in superconducting materials, we have investigated vortex states for the $s+id$-wave superconductor by Ginzburg-Landau simulations. The superconducting state is characterized by the competition between the two coupled condensates, and the interaction of the corresponding $s$-wave and $d$-wave superconducting order parameters.  We reveal how the variation in the relative strength of those condensates leads to the changes in their competition and changes in the ground state, so that the $s+id$ superconducting phase under magnetic field can exhibit three distinct states of vortex matter, i.e. the vortex lattice, skyrmionic vortex chains and the skyrmion lattice.  Vortex lattices occur when one of the component is absolutely dominant; skyrmionic vortex chains appear when one of the components is relatively dominant; the skyrmion lattices occur when two condensates have a matching strength.

In formation of such rich and peculiar vortical states we identified two key competitions.  One is the competition between different pairing symmetries of $s$- and $d$-wave condensates, resulting in the crucial characteristic of the vortex, where two condensates tend to have an opposite phase winding. Another is the segregative tendency of the $s+id$ and $s-id$ state, resulting in the formation of phase domain walls interconnecting the vortex cores in $s-$ and $d-$ condensates. We show that such walls enclosing domains of $\pi/2$ state difference between the condensates, surrounded by $-\pi/2$ state difference, carry topological charge and therefore have skyrmionic character. Due to their particular organizational geometry and their characteristic magnetic field distributions, each of the three characteristic states in $s+id$ superconductors can be distinctively observed experimentally, in e.g. scanning Hall and scanning SQUID experiments.

We emphasize that the skyrmionic vortex chains are a novel and unusual state, completely different from the pure skyrmion (coreless vortex) states.  The pure skyrmion states result from the segregative tendency between chiral states.  In such states, the composite vortex splits to form coreless vortex.  On the other hand, the skyrmionic vortex chains result also from the competing symmetries. Such symmetries ($s-$ and $d-$ condensates in our case) impose the opposite phase winding between two components, preventing the composite vortex from splitting its core.  Meanwhile, restriction of the topology in such case induces the coreless vortices in order to keeping the same phase winding for both condensates.  Finally, in the phase of skyrmionic vortex chains, the composite vortices are connected in a chain through coreless vortices, forming exotic vortex pattern and exhibiting skyrmionic topology.  Such skyrmionic vortex chain cannot be obtained trivially, by mere transformation from the coreless vortices, and has no analogy in the $s+is$ and $p+ip$ states.  As a result, the skyrmionic vortex chains provide a unique signature to distinguish $s+id$ state from $s+is$ and $p+ip$ states.  Moreover, they are a unique example in multi-component superconductors with competing orders where singular core vortices and coreless vortices combine to create a novel topological object.

The $s+id$ superconductivity is relevant for high-T$_c$ cuprates, e.g. YBCO and BSCCO, and iron-based superconductors, e.g.  Ba$_{1-x}$K$_x$Fe$_2$As$_2$.  The high-T$_c$ cuprates originally have a $d$-wave pairing symmetry,\cite{shenCuprateHighTcSuperconductors2008} but the existence of a subdominant $s$-wave symmetry has long been considered near inhomogeneous regions around defects and vortices, and in nanoengineered samples.  Recently observed nodeless superconducting gap in a CuO$_2$ monolayer on BSCCO is interpreted as an intrinsic $s$-wave state, \cite{zhongNodelessPairingSuperconducting2016, jiangNodelessHighSuperconductivity2018} opening another possibility for realizing the competition between the $s$-wave and the $d$-wave pairing symmetry.  That competition in the above systems is expected to be dependent on temperature (thus can be tunable). In the iron-based superconductor Ba$_{1-x}$K$_x$Fe$_2$As$_2$ $s+id$ state is expected according to the theoretical arguments, with competition between the condensates tunable by doping levels. Our results about the unique vortex states to be expected in $s+id$ state can therefore serve as its ubiquitous experimental proof in these and other superconducting materials of interest.

Very recently, it was shown that the second-order topological superconductor can be induced by the s+id state when brought in proximity to a Rashba spin-orbit coupled semiconductor. \cite{PhysRevLett.122.236401}  Topological superconductors may support Majorana zero modes in vortex cores, the control of which can enable quantum computations.  Our results here are a needed step towards understanding properties of vortices in such systems.

Finally, we remark that the Ginzburg-Landau simulations used in this paper are on the mean-field level, thus, neglecting fluctuations, which is appropriate for sufficiently low temperature.  Since the strength of the fluctuations grows with temperature, new phenomena such as vortical/skyrmionic lattice melting\cite{huFunctionPeakSpecific1997, smorgravVortexSublatticeMelting2005} may take place in the high temperature range, worth of further investigation.

\section*{Acknowledgments}
The authors acknowledge useful discussions with Prof. Yong-Ping Zhang. This research was supported by the National Natural Science Foundation of China under Grant No. 61571277 and No. 61771298. L.-F.Z. and M.V.M. acknowledge support from Research Foundation-Flanders (FWO-Vlaanderen).


\begin{thebibliography}{68}%
\makeatletter
\providecommand \@ifxundefined [1]{%
 \@ifx{#1\undefined}
}%
\providecommand \@ifnum [1]{%
 \ifnum #1\expandafter \@firstoftwo
 \else \expandafter \@secondoftwo
 \fi
}%
\providecommand \@ifx [1]{%
 \ifx #1\expandafter \@firstoftwo
 \else \expandafter \@secondoftwo
 \fi
}%
\providecommand \natexlab [1]{#1}%
\providecommand \enquote  [1]{``#1''}%
\providecommand \bibnamefont  [1]{#1}%
\providecommand \bibfnamefont [1]{#1}%
\providecommand \citenamefont [1]{#1}%
\providecommand \href@noop [0]{\@secondoftwo}%
\providecommand \href [0]{\begingroup \@sanitize@url \@href}%
\providecommand \@href[1]{\@@startlink{#1}\@@href}%
\providecommand \@@href[1]{\endgroup#1\@@endlink}%
\providecommand \@sanitize@url [0]{\catcode `\\12\catcode `\$12\catcode
  `\&12\catcode `\#12\catcode `\^12\catcode `\_12\catcode `\%12\relax}%
\providecommand \@@startlink[1]{}%
\providecommand \@@endlink[0]{}%
\providecommand \url  [0]{\begingroup\@sanitize@url \@url }%
\providecommand \@url [1]{\endgroup\@href {#1}{\urlprefix }}%
\providecommand \urlprefix  [0]{URL }%
\providecommand \Eprint [0]{\href }%
\providecommand \doibase [0]{http://dx.doi.org/}%
\providecommand \selectlanguage [0]{\@gobble}%
\providecommand \bibinfo  [0]{\@secondoftwo}%
\providecommand \bibfield  [0]{\@secondoftwo}%
\providecommand \translation [1]{[#1]}%
\providecommand \BibitemOpen [0]{}%
\providecommand \bibitemStop [0]{}%
\providecommand \bibitemNoStop [0]{.\EOS\space}%
\providecommand \EOS [0]{\spacefactor3000\relax}%
\providecommand \BibitemShut  [1]{\csname bibitem#1\endcsname}%
\let\auto@bib@innerbib\@empty
\bibitem [{\citenamefont {Milo{\v s}evi\'c}\ and\ \citenamefont
  {Perali}(2015)}]{milosevicEmergentPhenomenaMulticomponent2015}%
  \BibitemOpen
  \bibfield  {author} {\bibinfo {author} {\bibfnamefont {Milorad~V.}\
  \bibnamefont {Milo{\v s}evi\'c}}\ and\ \bibinfo {author} {\bibfnamefont
  {Andrea}\ \bibnamefont {Perali}},\ }\bibfield  {title} {\enquote {\bibinfo
  {title} {Emergent phenomena in multicomponent superconductivity: An
  introduction to the focus issue},}\ }\href {\doibase
  10.1088/0953-2048/28/6/060201} {\bibfield  {journal} {\bibinfo  {journal}
  {Supercond. Sci. Technol.}\ }\textbf {\bibinfo {volume} {28}},\ \bibinfo
  {pages} {060201} (\bibinfo {year} {2015})}\BibitemShut {NoStop}%
\bibitem [{\citenamefont {Sato}\ and\ \citenamefont
  {Ando}(2017)}]{satoTopologicalSuperconductorsReview2017}%
  \BibitemOpen
  \bibfield  {author} {\bibinfo {author} {\bibfnamefont {Masatoshi}\
  \bibnamefont {Sato}}\ and\ \bibinfo {author} {\bibfnamefont {Yoichi}\
  \bibnamefont {Ando}},\ }\bibfield  {title} {\enquote {\bibinfo {title}
  {Topological superconductors: a review},}\ }\href {\doibase
  10.1088/1361-6633/aa6ac7} {\bibfield  {journal} {\bibinfo  {journal} {Rep.
  Prog. Phys.}\ }\textbf {\bibinfo {volume} {80}},\ \bibinfo {pages} {076501}
  (\bibinfo {year} {2017})}\BibitemShut {NoStop}%
\bibitem [{\citenamefont {Mackenzie}\ and\ \citenamefont
  {Maeno}(2003)}]{mackenzieSuperconductivityMathrmSrMathrmRuO2003}%
  \BibitemOpen
  \bibfield  {author} {\bibinfo {author} {\bibfnamefont {Andrew~Peter}\
  \bibnamefont {Mackenzie}}\ and\ \bibinfo {author} {\bibfnamefont {Yoshiteru}\
  \bibnamefont {Maeno}},\ }\bibfield  {title} {\enquote {\bibinfo {title} {The
  superconductivity of {${\mathrm{Sr}}_{2}{\mathrm{RuO}}_{4}$} and the physics
  of spin-triplet pairing},}\ }\href {\doibase 10.1103/RevModPhys.75.657}
  {\bibfield  {journal} {\bibinfo  {journal} {Rev. Mod. Phys.}\ }\textbf
  {\bibinfo {volume} {75}},\ \bibinfo {pages} {657--712} (\bibinfo {year}
  {2003})}\BibitemShut {NoStop}%
\bibitem [{\citenamefont {Kallin}(2012)}]{kallinChiralPwaveOrder2012}%
  \BibitemOpen
  \bibfield  {author} {\bibinfo {author} {\bibfnamefont {Catherine}\
  \bibnamefont {Kallin}},\ }\bibfield  {title} {\enquote {\bibinfo {title}
  {Chiral p-wave order in {{Sr$_2$RuO$_4$}}},}\ }\href {\doibase
  10.1088/0034-4885/75/4/042501} {\bibfield  {journal} {\bibinfo  {journal}
  {Rep. Prog. Phys.}\ }\textbf {\bibinfo {volume} {75}},\ \bibinfo {pages}
  {042501} (\bibinfo {year} {2012})}\BibitemShut {NoStop}%
\bibitem [{\citenamefont {Nandkishore}\ \emph {et~al.}(2012)\citenamefont
  {Nandkishore}, \citenamefont {Levitov},\ and\ \citenamefont
  {Chubukov}}]{nandkishoreChiralSuperconductivityRepulsive2012}%
  \BibitemOpen
  \bibfield  {author} {\bibinfo {author} {\bibfnamefont {Rahul}\ \bibnamefont
  {Nandkishore}}, \bibinfo {author} {\bibfnamefont {L.~S.}\ \bibnamefont
  {Levitov}}, \ and\ \bibinfo {author} {\bibfnamefont {A.~V.}\ \bibnamefont
  {Chubukov}},\ }\bibfield  {title} {\enquote {\bibinfo {title} {Chiral
  superconductivity from repulsive interactions in doped graphene},}\ }\href
  {\doibase 10.1038/nphys2208} {\bibfield  {journal} {\bibinfo  {journal}
  {Nature Physics}\ }\textbf {\bibinfo {volume} {8}},\ \bibinfo {pages}
  {158--163} (\bibinfo {year} {2012})}\BibitemShut {NoStop}%
\bibitem [{\citenamefont {{Black-Schaffer}}\ and\ \citenamefont
  {Honerkamp}(2014)}]{black-schafferChiralWaveSuperconductivity2014}%
  \BibitemOpen
  \bibfield  {author} {\bibinfo {author} {\bibfnamefont {Annica~M.}\
  \bibnamefont {{Black-Schaffer}}}\ and\ \bibinfo {author} {\bibfnamefont
  {Carsten}\ \bibnamefont {Honerkamp}},\ }\bibfield  {title} {\enquote
  {\bibinfo {title} {Chiral d-wave superconductivity in doped graphene},}\
  }\href {\doibase 10.1088/0953-8984/26/42/423201} {\bibfield  {journal}
  {\bibinfo  {journal} {J. Phys.: Condens. Matter}\ }\textbf {\bibinfo {volume}
  {26}},\ \bibinfo {pages} {423201} (\bibinfo {year} {2014})}\BibitemShut
  {NoStop}%
\bibitem [{\citenamefont {Fischer}\ \emph {et~al.}(2014)\citenamefont
  {Fischer}, \citenamefont {Neupert}, \citenamefont {Platt}, \citenamefont
  {Schnyder}, \citenamefont {Hanke}, \citenamefont {Goryo}, \citenamefont
  {Thomale},\ and\ \citenamefont
  {Sigrist}}]{fischerChiralWaveSuperconductivity2014}%
  \BibitemOpen
  \bibfield  {author} {\bibinfo {author} {\bibfnamefont {Mark~H.}\ \bibnamefont
  {Fischer}}, \bibinfo {author} {\bibfnamefont {Titus}\ \bibnamefont
  {Neupert}}, \bibinfo {author} {\bibfnamefont {Christian}\ \bibnamefont
  {Platt}}, \bibinfo {author} {\bibfnamefont {Andreas~P.}\ \bibnamefont
  {Schnyder}}, \bibinfo {author} {\bibfnamefont {Werner}\ \bibnamefont
  {Hanke}}, \bibinfo {author} {\bibfnamefont {Jun}\ \bibnamefont {Goryo}},
  \bibinfo {author} {\bibfnamefont {Ronny}\ \bibnamefont {Thomale}}, \ and\
  \bibinfo {author} {\bibfnamefont {Manfred}\ \bibnamefont {Sigrist}},\
  }\bibfield  {title} {\enquote {\bibinfo {title} {Chiral $d$-wave
  superconductivity in srptas},}\ }\href {\doibase 10.1103/PhysRevB.89.020509}
  {\bibfield  {journal} {\bibinfo  {journal} {Phys. Rev. B}\ }\textbf {\bibinfo
  {volume} {89}},\ \bibinfo {pages} {020509(R)} (\bibinfo {year}
  {2014})}\BibitemShut {NoStop}%
\bibitem [{\citenamefont
  {Tanaka}(2015)}]{tanakaMulticomponentSuperconductivityBased2015}%
  \BibitemOpen
  \bibfield  {author} {\bibinfo {author} {\bibfnamefont {Y.}~\bibnamefont
  {Tanaka}},\ }\bibfield  {title} {\enquote {\bibinfo {title} {Multicomponent
  superconductivity based on multiband superconductors},}\ }\href {\doibase
  10.1088/0953-2048/28/3/034002} {\bibfield  {journal} {\bibinfo  {journal}
  {Supercond. Sci. Technol.}\ }\textbf {\bibinfo {volume} {28}},\ \bibinfo
  {pages} {034002} (\bibinfo {year} {2015})}\BibitemShut {NoStop}%
\bibitem [{\citenamefont {Lin}(2014)}]{linGroundStateCollective2014}%
  \BibitemOpen
  \bibfield  {author} {\bibinfo {author} {\bibfnamefont {Shi-Zeng}\
  \bibnamefont {Lin}},\ }\bibfield  {title} {\enquote {\bibinfo {title} {Ground
  state, collective mode, phase soliton and vortex in multiband
  superconductors},}\ }\href {\doibase 10.1088/0953-8984/26/49/493202}
  {\bibfield  {journal} {\bibinfo  {journal} {J. Phys.: Condens. Matter}\
  }\textbf {\bibinfo {volume} {26}},\ \bibinfo {pages} {493202} (\bibinfo
  {year} {2014})}\BibitemShut {NoStop}%
\bibitem [{\citenamefont {Babaev}\ and\ \citenamefont
  {Speight}(2005)}]{babaevSemiMeissnerStateNeither2005}%
  \BibitemOpen
  \bibfield  {author} {\bibinfo {author} {\bibfnamefont {Egor}\ \bibnamefont
  {Babaev}}\ and\ \bibinfo {author} {\bibfnamefont {Martin}\ \bibnamefont
  {Speight}},\ }\bibfield  {title} {\enquote {\bibinfo {title} {Semi-meissner
  state and neither type-{I} nor type-{II} superconductivity in multicomponent
  superconductors},}\ }\href {\doibase 10.1103/PhysRevB.72.180502} {\bibfield
  {journal} {\bibinfo  {journal} {Phys. Rev. B}\ }\textbf {\bibinfo {volume}
  {72}},\ \bibinfo {pages} {180502(R)} (\bibinfo {year} {2005})}\BibitemShut
  {NoStop}%
\bibitem [{\citenamefont {Lin}\ and\ \citenamefont
  {Hu}(2011)}]{linVortexStatesPhase2011}%
  \BibitemOpen
  \bibfield  {author} {\bibinfo {author} {\bibfnamefont {Shi-Zeng}\
  \bibnamefont {Lin}}\ and\ \bibinfo {author} {\bibfnamefont {Xiao}\
  \bibnamefont {Hu}},\ }\bibfield  {title} {\enquote {\bibinfo {title} {Vortex
  states and the phase diagram of a multiple-component ginzburg-landau theory
  with competing repulsive and attractive vortex interactions},}\ }\href
  {\doibase 10.1103/PhysRevB.84.214505} {\bibfield  {journal} {\bibinfo
  {journal} {Phys. Rev. B}\ }\textbf {\bibinfo {volume} {84}},\ \bibinfo
  {pages} {214505} (\bibinfo {year} {2011})}\BibitemShut {NoStop}%
\bibitem [{\citenamefont {Xu}\ \emph {et~al.}(2014)\citenamefont {Xu},
  \citenamefont {Fangohr}, \citenamefont {Gu}, \citenamefont {Chen},
  \citenamefont {Wang}, \citenamefont {Zhou}, \citenamefont {Shi},\ and\
  \citenamefont {Dou}}]{xuSimulationPhaseDiagram2014}%
  \BibitemOpen
  \bibfield  {author} {\bibinfo {author} {\bibfnamefont {X.~B.}\ \bibnamefont
  {Xu}}, \bibinfo {author} {\bibfnamefont {H.}~\bibnamefont {Fangohr}},
  \bibinfo {author} {\bibfnamefont {M.}~\bibnamefont {Gu}}, \bibinfo {author}
  {\bibfnamefont {W.}~\bibnamefont {Chen}}, \bibinfo {author} {\bibfnamefont
  {Z.~H.}\ \bibnamefont {Wang}}, \bibinfo {author} {\bibfnamefont
  {F.}~\bibnamefont {Zhou}}, \bibinfo {author} {\bibfnamefont {D.~Q.}\
  \bibnamefont {Shi}}, \ and\ \bibinfo {author} {\bibfnamefont {S.~X.}\
  \bibnamefont {Dou}},\ }\bibfield  {title} {\enquote {\bibinfo {title}
  {Simulation of the phase diagram of magnetic vortices in two-dimensional
  superconductors: Evidence for vortex chain formation},}\ }\href {\doibase
  10.1088/0953-8984/26/11/115702} {\bibfield  {journal} {\bibinfo  {journal}
  {J. Phys.: Condens. Matter}\ }\textbf {\bibinfo {volume} {26}},\ \bibinfo
  {pages} {115702} (\bibinfo {year} {2014})}\BibitemShut {NoStop}%
\bibitem [{\citenamefont
  {Tanaka}(2001)}]{tanakaSolitonTwoBandSuperconductor2001}%
  \BibitemOpen
  \bibfield  {author} {\bibinfo {author} {\bibfnamefont {Y.}~\bibnamefont
  {Tanaka}},\ }\bibfield  {title} {\enquote {\bibinfo {title} {Soliton in
  {{Two}}-{{Band Superconductor}}},}\ }\href {\doibase
  10.1103/PhysRevLett.88.017002} {\bibfield  {journal} {\bibinfo  {journal}
  {Phys. Rev. Lett.}\ }\textbf {\bibinfo {volume} {88}},\ \bibinfo {pages}
  {017002} (\bibinfo {year} {2001})}\BibitemShut {NoStop}%
\bibitem [{\citenamefont {Lin}\ and\ \citenamefont
  {Hu}(2012)}]{linPhaseSolitonsMultiband2012}%
  \BibitemOpen
  \bibfield  {author} {\bibinfo {author} {\bibfnamefont {Shi-Zeng}\
  \bibnamefont {Lin}}\ and\ \bibinfo {author} {\bibfnamefont {Xiao}\
  \bibnamefont {Hu}},\ }\bibfield  {title} {\enquote {\bibinfo {title} {Phase
  solitons in multi-band superconductors with and without time-reversal
  symmetry},}\ }\href {\doibase 10.1088/1367-2630/14/6/063021} {\bibfield
  {journal} {\bibinfo  {journal} {New J. Phys.}\ }\textbf {\bibinfo {volume}
  {14}},\ \bibinfo {pages} {063021} (\bibinfo {year} {2012})}\BibitemShut
  {NoStop}%
\bibitem [{\citenamefont {Matsumoto}\ and\ \citenamefont
  {Sigrist}(1999)}]{matsumotoQuasiparticleStatesSurface1999}%
  \BibitemOpen
  \bibfield  {author} {\bibinfo {author} {\bibfnamefont {Masashige}\
  \bibnamefont {Matsumoto}}\ and\ \bibinfo {author} {\bibfnamefont {Manfred}\
  \bibnamefont {Sigrist}},\ }\bibfield  {title} {\enquote {\bibinfo {title}
  {Quasiparticle {{States}} near the {{Surface}} and the {{Domain Wall}} in a
  ${p}_{x} \pm i{p}_{y}$-{{Wave Superconductor}}},}\ }\href {\doibase
  10.1143/JPSJ.68.994} {\bibfield  {journal} {\bibinfo  {journal} {J. Phys.
  Soc. Jpn.}\ }\textbf {\bibinfo {volume} {68}},\ \bibinfo {pages} {994--1007}
  (\bibinfo {year} {1999})}\BibitemShut {NoStop}%
\bibitem [{\citenamefont {Garaud}\ and\ \citenamefont
  {Babaev}(2014)}]{garaudDomainWallsTheir2014}%
  \BibitemOpen
  \bibfield  {author} {\bibinfo {author} {\bibfnamefont {Julien}\ \bibnamefont
  {Garaud}}\ and\ \bibinfo {author} {\bibfnamefont {Egor}\ \bibnamefont
  {Babaev}},\ }\bibfield  {title} {\enquote {\bibinfo {title} {Domain walls and
  their experimental signatures in $s+is$ superconductors},}\ }\href {\doibase
  10.1103/PhysRevLett.112.017003} {\bibfield  {journal} {\bibinfo  {journal}
  {Phys. Rev. Lett.}\ }\textbf {\bibinfo {volume} {112}},\ \bibinfo {pages}
  {017003} (\bibinfo {year} {2014})}\BibitemShut {NoStop}%
\bibitem [{\citenamefont {Zhang}\ \emph {et~al.}(2017)\citenamefont {Zhang},
  \citenamefont {Covaci},\ and\ \citenamefont {Milo{\v
  s}evi\'c}}]{zhangTopologicalPhaseTransitions2017}%
  \BibitemOpen
  \bibfield  {author} {\bibinfo {author} {\bibfnamefont {L.-F.}\ \bibnamefont
  {Zhang}}, \bibinfo {author} {\bibfnamefont {L.}~\bibnamefont {Covaci}}, \
  and\ \bibinfo {author} {\bibfnamefont {M.~V.}\ \bibnamefont {Milo{\v
  s}evi\'c}},\ }\bibfield  {title} {\enquote {\bibinfo {title} {Topological
  phase transitions in small mesoscopic chiral $p$-wave superconductors},}\
  }\href {\doibase 10.1103/PhysRevB.96.224512} {\bibfield  {journal} {\bibinfo
  {journal} {Phys. Rev. B}\ }\textbf {\bibinfo {volume} {96}},\ \bibinfo
  {pages} {224512} (\bibinfo {year} {2017})}\BibitemShut {NoStop}%
\bibitem [{\citenamefont {Babaev}\ \emph {et~al.}(2002)\citenamefont {Babaev},
  \citenamefont {Faddeev},\ and\ \citenamefont
  {Niemi}}]{babaevHiddenSymmetryKnot2002}%
  \BibitemOpen
  \bibfield  {author} {\bibinfo {author} {\bibfnamefont {Egor}\ \bibnamefont
  {Babaev}}, \bibinfo {author} {\bibfnamefont {Ludvig~D.}\ \bibnamefont
  {Faddeev}}, \ and\ \bibinfo {author} {\bibfnamefont {Antti~J.}\ \bibnamefont
  {Niemi}},\ }\bibfield  {title} {\enquote {\bibinfo {title} {Hidden symmetry
  and knot solitons in a charged two-condensate {Bose} system},}\ }\href
  {\doibase 10.1103/PhysRevB.65.100512} {\bibfield  {journal} {\bibinfo
  {journal} {Phys. Rev. B}\ }\textbf {\bibinfo {volume} {65}},\ \bibinfo
  {pages} {100512(R)} (\bibinfo {year} {2002})}\BibitemShut {NoStop}%
\bibitem [{\citenamefont {Zhang}\ \emph {et~al.}(2016)\citenamefont {Zhang},
  \citenamefont {Becerra}, \citenamefont {Covaci},\ and\ \citenamefont {Milo{\v
  s}evi\'c}}]{zhang_electronic_2016}%
  \BibitemOpen
  \bibfield  {author} {\bibinfo {author} {\bibfnamefont {L.-F.}\ \bibnamefont
  {Zhang}}, \bibinfo {author} {\bibfnamefont {V.~F.}\ \bibnamefont {Becerra}},
  \bibinfo {author} {\bibfnamefont {L.}~\bibnamefont {Covaci}}, \ and\ \bibinfo
  {author} {\bibfnamefont {M.~V.}\ \bibnamefont {Milo{\v s}evi\'c}},\
  }\bibfield  {title} {\enquote {\bibinfo {title} {Electronic properties of
  emergent topological defects in chiral $p$-wave superconductivity},}\ }\href
  {\doibase 10.1103/PhysRevB.94.024520} {\bibfield  {journal} {\bibinfo
  {journal} {Phys. Rev. B}\ }\textbf {\bibinfo {volume} {94}},\ \bibinfo
  {pages} {024520} (\bibinfo {year} {2016})}\BibitemShut {NoStop}%
\bibitem [{\citenamefont {Garaud}\ \emph
  {et~al.}(2016{\natexlab{a}})\citenamefont {Garaud}, \citenamefont {Babaev},
  \citenamefont {Bojesen},\ and\ \citenamefont
  {Sudb\o{}}}]{garaudLatticesDoublequantaVortices2016}%
  \BibitemOpen
  \bibfield  {author} {\bibinfo {author} {\bibfnamefont {Julien}\ \bibnamefont
  {Garaud}}, \bibinfo {author} {\bibfnamefont {Egor}\ \bibnamefont {Babaev}},
  \bibinfo {author} {\bibfnamefont {Troels~Arnfred}\ \bibnamefont {Bojesen}}, \
  and\ \bibinfo {author} {\bibfnamefont {Asle}\ \bibnamefont {Sudb\o{}}},\
  }\bibfield  {title} {\enquote {\bibinfo {title} {Lattices of double-quanta
  vortices and chirality inversion in ${p}_{x}+i{p}_{y}$ superconductors},}\
  }\href {\doibase 10.1103/PhysRevB.94.104509} {\bibfield  {journal} {\bibinfo
  {journal} {Phys. Rev. B}\ }\textbf {\bibinfo {volume} {94}},\ \bibinfo
  {pages} {104509} (\bibinfo {year} {2016}{\natexlab{a}})}\BibitemShut
  {NoStop}%
\bibitem [{\citenamefont {Garaud}\ and\ \citenamefont
  {Babaev}(2015{\natexlab{a}})}]{garaudPropertiesSkyrmionsMultiquanta2015}%
  \BibitemOpen
  \bibfield  {author} {\bibinfo {author} {\bibfnamefont {Julien}\ \bibnamefont
  {Garaud}}\ and\ \bibinfo {author} {\bibfnamefont {Egor}\ \bibnamefont
  {Babaev}},\ }\bibfield  {title} {\enquote {\bibinfo {title} {Properties of
  skyrmions and multi-quanta vortices in chiral p-wave superconductors},}\
  }\href {\doibase 10.1038/srep17540} {\bibfield  {journal} {\bibinfo
  {journal} {Scientific Reports}\ }\textbf {\bibinfo {volume} {5}},\ \bibinfo
  {pages} {17540} (\bibinfo {year} {2015}{\natexlab{a}})}\BibitemShut {NoStop}%
\bibitem [{\citenamefont {Fern\'andez~Becerra}\ \emph
  {et~al.}(2016)\citenamefont {Fern\'andez~Becerra}, \citenamefont {Sardella},
  \citenamefont {Peeters},\ and\ \citenamefont {Milo\ifmmode \check{s}\else
  \v{s}\fi{}evi\ifmmode~\acute{c}\else
  \'{c}\fi{}}}]{fernandezbecerraVorticalSkyrmionicStates2016}%
  \BibitemOpen
  \bibfield  {author} {\bibinfo {author} {\bibfnamefont {V.}~\bibnamefont
  {Fern\'andez~Becerra}}, \bibinfo {author} {\bibfnamefont {E.}~\bibnamefont
  {Sardella}}, \bibinfo {author} {\bibfnamefont {F.~M.}\ \bibnamefont
  {Peeters}}, \ and\ \bibinfo {author} {\bibfnamefont {M.~V.}\ \bibnamefont
  {Milo\ifmmode \check{s}\else \v{s}\fi{}evi\ifmmode~\acute{c}\else
  \'{c}\fi{}}},\ }\bibfield  {title} {\enquote {\bibinfo {title} {Vortical
  versus skyrmionic states in mesoscopic $p$-wave superconductors},}\ }\href
  {\doibase 10.1103/PhysRevB.93.014518} {\bibfield  {journal} {\bibinfo
  {journal} {Phys. Rev. B}\ }\textbf {\bibinfo {volume} {93}},\ \bibinfo
  {pages} {014518} (\bibinfo {year} {2016})}\BibitemShut {NoStop}%
\bibitem [{\citenamefont {Fern\'andez~Becerra}\ and\ \citenamefont
  {Milo\ifmmode \check{s}\else \v{s}\fi{}evi\ifmmode~\acute{c}\else
  \'{c}\fi{}}(2017)}]{fernandezbecerraDynamicsSkyrmionsEdge2017}%
  \BibitemOpen
  \bibfield  {author} {\bibinfo {author} {\bibfnamefont {V.}~\bibnamefont
  {Fern\'andez~Becerra}}\ and\ \bibinfo {author} {\bibfnamefont {M.~V.}\
  \bibnamefont {Milo\ifmmode \check{s}\else
  \v{s}\fi{}evi\ifmmode~\acute{c}\else \'{c}\fi{}}},\ }\bibfield  {title}
  {\enquote {\bibinfo {title} {Dynamics of skyrmions and edge states in the
  resistive regime of mesoscopic p-wave superconductors},}\ }\href {\doibase
  10.1016/j.physc.2016.07.002} {\bibfield  {journal} {\bibinfo  {journal}
  {Physica C}\ }\textbf {\bibinfo {volume} {533}},\ \bibinfo {pages} {91--95}
  (\bibinfo {year} {2017})}\BibitemShut {NoStop}%
\bibitem [{\citenamefont {Garaud}\ \emph {et~al.}(2013)\citenamefont {Garaud},
  \citenamefont {Carlstr\"om}, \citenamefont {Babaev},\ and\ \citenamefont
  {Speight}}]{garaudChiralMathbbCPSkyrmions2013}%
  \BibitemOpen
  \bibfield  {author} {\bibinfo {author} {\bibfnamefont {Julien}\ \bibnamefont
  {Garaud}}, \bibinfo {author} {\bibfnamefont {Johan}\ \bibnamefont
  {Carlstr\"om}}, \bibinfo {author} {\bibfnamefont {Egor}\ \bibnamefont
  {Babaev}}, \ and\ \bibinfo {author} {\bibfnamefont {Martin}\ \bibnamefont
  {Speight}},\ }\bibfield  {title} {\enquote {\bibinfo {title} {Chiral
  $\mathbb{C}{P}^{2}$ skyrmions in three-band superconductors},}\ }\href
  {\doibase 10.1103/PhysRevB.87.014507} {\bibfield  {journal} {\bibinfo
  {journal} {Phys. Rev. B}\ }\textbf {\bibinfo {volume} {87}},\ \bibinfo
  {pages} {014507} (\bibinfo {year} {2013})}\BibitemShut {NoStop}%
\bibitem [{\citenamefont {Zyuzin}\ \emph {et~al.}(2017)\citenamefont {Zyuzin},
  \citenamefont {Garaud},\ and\ \citenamefont
  {Babaev}}]{zyuzinNematicSkyrmionsOddParity2017}%
  \BibitemOpen
  \bibfield  {author} {\bibinfo {author} {\bibfnamefont {A.~A.}\ \bibnamefont
  {Zyuzin}}, \bibinfo {author} {\bibfnamefont {Julien}\ \bibnamefont {Garaud}},
  \ and\ \bibinfo {author} {\bibfnamefont {Egor}\ \bibnamefont {Babaev}},\
  }\bibfield  {title} {\enquote {\bibinfo {title} {Nematic skyrmions in
  odd-parity superconductors},}\ }\href {\doibase
  10.1103/PhysRevLett.119.167001} {\bibfield  {journal} {\bibinfo  {journal}
  {Phys. Rev. Lett.}\ }\textbf {\bibinfo {volume} {119}},\ \bibinfo {pages}
  {167001} (\bibinfo {year} {2017})}\BibitemShut {NoStop}%
\bibitem [{\citenamefont {Ho}(1978)}]{hoCorelessVorticesSuperfluid1978}%
  \BibitemOpen
  \bibfield  {author} {\bibinfo {author} {\bibfnamefont {Tin-Lun}\ \bibnamefont
  {Ho}},\ }\bibfield  {title} {\enquote {\bibinfo {title} {Coreless vortices in
  superfluid $^{3}\mathrm{He}$-{$A$}: Topological structure, nucleation, and
  the screening effect},}\ }\href {\doibase 10.1103/PhysRevB.18.1144}
  {\bibfield  {journal} {\bibinfo  {journal} {Phys. Rev. B}\ }\textbf {\bibinfo
  {volume} {18}},\ \bibinfo {pages} {1144--1153} (\bibinfo {year}
  {1978})}\BibitemShut {NoStop}%
\bibitem [{\citenamefont {Mermin}\ and\ \citenamefont
  {Ho}(1976)}]{merminCirculationAngularMomentum1976}%
  \BibitemOpen
  \bibfield  {author} {\bibinfo {author} {\bibfnamefont {N.~D.}\ \bibnamefont
  {Mermin}}\ and\ \bibinfo {author} {\bibfnamefont {Tin-Lun}\ \bibnamefont
  {Ho}},\ }\bibfield  {title} {\enquote {\bibinfo {title} {Circulation and
  angular momentum in the {$A$} phase of superfluid {Helium-3}},}\ }\href
  {\doibase 10.1103/PhysRevLett.36.594} {\bibfield  {journal} {\bibinfo
  {journal} {Phys. Rev. Lett.}\ }\textbf {\bibinfo {volume} {36}},\ \bibinfo
  {pages} {594--597} (\bibinfo {year} {1976})}\BibitemShut {NoStop}%
\bibitem [{\citenamefont {Mizushima}\ \emph {et~al.}(2002)\citenamefont
  {Mizushima}, \citenamefont {Machida},\ and\ \citenamefont
  {Kita}}]{mizushimaMerminHoVortexFerromagnetic2002}%
  \BibitemOpen
  \bibfield  {author} {\bibinfo {author} {\bibfnamefont {T.}~\bibnamefont
  {Mizushima}}, \bibinfo {author} {\bibfnamefont {K.}~\bibnamefont {Machida}},
  \ and\ \bibinfo {author} {\bibfnamefont {T.}~\bibnamefont {Kita}},\
  }\bibfield  {title} {\enquote {\bibinfo {title} {{Mermin-Ho Vortex in
  Ferromagnetic Spinor Bose-Einstein Condensates}},}\ }\href {\doibase
  10.1103/PhysRevLett.89.030401} {\bibfield  {journal} {\bibinfo  {journal}
  {Phys. Rev. Lett.}\ }\textbf {\bibinfo {volume} {89}},\ \bibinfo {pages}
  {030401} (\bibinfo {year} {2002})}\BibitemShut {NoStop}%
\bibitem [{\citenamefont {Anderson}\ and\ \citenamefont
  {Toulouse}(1977)}]{andersonPhaseSlippageVortex1977}%
  \BibitemOpen
  \bibfield  {author} {\bibinfo {author} {\bibfnamefont {P.~W.}\ \bibnamefont
  {Anderson}}\ and\ \bibinfo {author} {\bibfnamefont {G.}~\bibnamefont
  {Toulouse}},\ }\bibfield  {title} {\enquote {\bibinfo {title} {Phase slippage
  without vortex cores: Vortex textures in superfluid $^{3}\mathrm{He}$},}\
  }\href {\doibase 10.1103/PhysRevLett.38.508} {\bibfield  {journal} {\bibinfo
  {journal} {Phys. Rev. Lett.}\ }\textbf {\bibinfo {volume} {38}},\ \bibinfo
  {pages} {508--511} (\bibinfo {year} {1977})}\BibitemShut {NoStop}%
\bibitem [{\citenamefont {Ho}(1998)}]{PhysRevLett.81.742}%
  \BibitemOpen
  \bibfield  {author} {\bibinfo {author} {\bibfnamefont {Tin-Lun}\ \bibnamefont
  {Ho}},\ }\bibfield  {title} {\enquote {\bibinfo {title} {{Spinor Bose
  Condensates in Optical Traps}},}\ }\href {\doibase
  10.1103/PhysRevLett.81.742} {\bibfield  {journal} {\bibinfo  {journal} {Phys.
  Rev. Lett.}\ }\textbf {\bibinfo {volume} {81}},\ \bibinfo {pages} {742--745}
  (\bibinfo {year} {1998})}\BibitemShut {NoStop}%
\bibitem [{\citenamefont {Leanhardt}\ \emph {et~al.}(2003)\citenamefont
  {Leanhardt}, \citenamefont {Shin}, \citenamefont {Kielpinski}, \citenamefont
  {Pritchard},\ and\ \citenamefont {Ketterle}}]{PhysRevLett.90.140403}%
  \BibitemOpen
  \bibfield  {author} {\bibinfo {author} {\bibfnamefont {A.~E.}\ \bibnamefont
  {Leanhardt}}, \bibinfo {author} {\bibfnamefont {Y.}~\bibnamefont {Shin}},
  \bibinfo {author} {\bibfnamefont {D.}~\bibnamefont {Kielpinski}}, \bibinfo
  {author} {\bibfnamefont {D.~E.}\ \bibnamefont {Pritchard}}, \ and\ \bibinfo
  {author} {\bibfnamefont {W.}~\bibnamefont {Ketterle}},\ }\bibfield  {title}
  {\enquote {\bibinfo {title} {{Coreless Vortex Formation in a Spinor
  Bose-Einstein Condensate}},}\ }\href {\doibase 10.1103/PhysRevLett.90.140403}
  {\bibfield  {journal} {\bibinfo  {journal} {Phys. Rev. Lett.}\ }\textbf
  {\bibinfo {volume} {90}},\ \bibinfo {pages} {140403} (\bibinfo {year}
  {2003})}\BibitemShut {NoStop}%
\bibitem [{\citenamefont {Catelani}\ and\ \citenamefont
  {Yuzbashyan}(2010)}]{PhysRevA.81.033629}%
  \BibitemOpen
  \bibfield  {author} {\bibinfo {author} {\bibfnamefont {G.}~\bibnamefont
  {Catelani}}\ and\ \bibinfo {author} {\bibfnamefont {E.~A.}\ \bibnamefont
  {Yuzbashyan}},\ }\bibfield  {title} {\enquote {\bibinfo {title} {{Coreless
  vorticity in multicomponent Bose and Fermi superfluids}},}\ }\href {\doibase
  10.1103/PhysRevA.81.033629} {\bibfield  {journal} {\bibinfo  {journal} {Phys.
  Rev. A}\ }\textbf {\bibinfo {volume} {81}},\ \bibinfo {pages} {033629}
  (\bibinfo {year} {2010})}\BibitemShut {NoStop}%
\bibitem [{\citenamefont {Mizushima}\ \emph {et~al.}(2004)\citenamefont
  {Mizushima}, \citenamefont {Kobayashi},\ and\ \citenamefont
  {Machida}}]{mizushimaCorelessSingularVortex2004}%
  \BibitemOpen
  \bibfield  {author} {\bibinfo {author} {\bibfnamefont {Takeshi}\ \bibnamefont
  {Mizushima}}, \bibinfo {author} {\bibfnamefont {Naoko}\ \bibnamefont
  {Kobayashi}}, \ and\ \bibinfo {author} {\bibfnamefont {Kazushige}\
  \bibnamefont {Machida}},\ }\bibfield  {title} {\enquote {\bibinfo {title}
  {Coreless and singular vortex lattices in rotating spinor {Bose-Einstein}
  condensates},}\ }\href {\doibase 10.1103/PhysRevA.70.043613} {\bibfield
  {journal} {\bibinfo  {journal} {Phys. Rev. A}\ }\textbf {\bibinfo {volume}
  {70}},\ \bibinfo {pages} {043613} (\bibinfo {year} {2004})}\BibitemShut
  {NoStop}%
\bibitem [{\citenamefont {Choi}\ \emph {et~al.}(2012)\citenamefont {Choi},
  \citenamefont {Kwon},\ and\ \citenamefont
  {Shin}}]{choiObservationTopologicallyStable2012}%
  \BibitemOpen
  \bibfield  {author} {\bibinfo {author} {\bibfnamefont {Jae-yoon}\
  \bibnamefont {Choi}}, \bibinfo {author} {\bibfnamefont {Woo~Jin}\
  \bibnamefont {Kwon}}, \ and\ \bibinfo {author} {\bibfnamefont {Yong-il}\
  \bibnamefont {Shin}},\ }\bibfield  {title} {\enquote {\bibinfo {title}
  {{Observation of Topologically Stable 2D Skyrmions in an Antiferromagnetic
  Spinor Bose-Einstein Condensate}},}\ }\href {\doibase
  10.1103/PhysRevLett.108.035301} {\bibfield  {journal} {\bibinfo  {journal}
  {Phys. Rev. Lett.}\ }\textbf {\bibinfo {volume} {108}},\ \bibinfo {pages}
  {035301} (\bibinfo {year} {2012})}\BibitemShut {NoStop}%
\bibitem [{\citenamefont {Orlova}\ \emph {et~al.}(2016)\citenamefont {Orlova},
  \citenamefont {Kuopanportti},\ and\ \citenamefont {Milo\ifmmode
  \check{s}\else \v{s}\fi{}evi\ifmmode~\acute{c}\else
  \'{c}\fi{}}}]{orlovaSkyrmionicVortexLattices2016}%
  \BibitemOpen
  \bibfield  {author} {\bibinfo {author} {\bibfnamefont {Natalia~V.}\
  \bibnamefont {Orlova}}, \bibinfo {author} {\bibfnamefont {Pekko}\
  \bibnamefont {Kuopanportti}}, \ and\ \bibinfo {author} {\bibfnamefont
  {Milorad~V.}\ \bibnamefont {Milo\ifmmode \check{s}\else
  \v{s}\fi{}evi\ifmmode~\acute{c}\else \'{c}\fi{}}},\ }\bibfield  {title}
  {\enquote {\bibinfo {title} {Skyrmionic vortex lattices in coherently coupled
  three-component {Bose-Einstein} condensates},}\ }\href {\doibase
  10.1103/PhysRevA.94.023617} {\bibfield  {journal} {\bibinfo  {journal} {Phys.
  Rev. A}\ }\textbf {\bibinfo {volume} {94}},\ \bibinfo {pages} {023617}
  (\bibinfo {year} {2016})}\BibitemShut {NoStop}%
\bibitem [{\citenamefont {Kuopanportti}\ \emph {et~al.}(2015)\citenamefont
  {Kuopanportti}, \citenamefont {Orlova},\ and\ \citenamefont {Milo\ifmmode
  \check{s}\else \v{s}\fi{}evi\ifmmode~\acute{c}\else
  \'{c}\fi{}}}]{PhysRevA.91.043605}%
  \BibitemOpen
  \bibfield  {author} {\bibinfo {author} {\bibfnamefont {Pekko}\ \bibnamefont
  {Kuopanportti}}, \bibinfo {author} {\bibfnamefont {Natalia~V.}\ \bibnamefont
  {Orlova}}, \ and\ \bibinfo {author} {\bibfnamefont {Milorad~V.}\ \bibnamefont
  {Milo\ifmmode \check{s}\else \v{s}\fi{}evi\ifmmode~\acute{c}\else
  \'{c}\fi{}}},\ }\bibfield  {title} {\enquote {\bibinfo {title} {Ground-state
  multiquantum vortices in rotating two-species superfluids},}\ }\href
  {\doibase 10.1103/PhysRevA.91.043605} {\bibfield  {journal} {\bibinfo
  {journal} {Phys. Rev. A}\ }\textbf {\bibinfo {volume} {91}},\ \bibinfo
  {pages} {043605} (\bibinfo {year} {2015})}\BibitemShut {NoStop}%
\bibitem [{\citenamefont {Yang}\ \emph {et~al.}(2008)\citenamefont {Yang},
  \citenamefont {Wu}, \citenamefont {Zhang},\ and\ \citenamefont
  {Feng}}]{PhysRevA.77.033621}%
  \BibitemOpen
  \bibfield  {author} {\bibinfo {author} {\bibfnamefont {Shi-Jie}\ \bibnamefont
  {Yang}}, \bibinfo {author} {\bibfnamefont {Quan-Sheng}\ \bibnamefont {Wu}},
  \bibinfo {author} {\bibfnamefont {Sheng-Nan}\ \bibnamefont {Zhang}}, \ and\
  \bibinfo {author} {\bibfnamefont {Shiping}\ \bibnamefont {Feng}},\ }\bibfield
   {title} {\enquote {\bibinfo {title} {Giant vortex and skyrmion in a rotating
  two-species {Bose-Einstein} condensate},}\ }\href {\doibase
  10.1103/PhysRevA.77.033621} {\bibfield  {journal} {\bibinfo  {journal} {Phys.
  Rev. A}\ }\textbf {\bibinfo {volume} {77}},\ \bibinfo {pages} {033621}
  (\bibinfo {year} {2008})}\BibitemShut {NoStop}%
\bibitem [{\citenamefont {Lovegrove}\ \emph {et~al.}(2014)\citenamefont
  {Lovegrove}, \citenamefont {Borgh},\ and\ \citenamefont
  {Ruostekoski}}]{PhysRevLett.112.075301}%
  \BibitemOpen
  \bibfield  {author} {\bibinfo {author} {\bibfnamefont {Justin}\ \bibnamefont
  {Lovegrove}}, \bibinfo {author} {\bibfnamefont {Magnus~O.}\ \bibnamefont
  {Borgh}}, \ and\ \bibinfo {author} {\bibfnamefont {Janne}\ \bibnamefont
  {Ruostekoski}},\ }\bibfield  {title} {\enquote {\bibinfo {title} {{Energetic
  Stability of Coreless Vortices in Spin-1 Bose-Einstein Condensates with
  Conserved Magnetization}},}\ }\href {\doibase 10.1103/PhysRevLett.112.075301}
  {\bibfield  {journal} {\bibinfo  {journal} {Phys. Rev. Lett.}\ }\textbf
  {\bibinfo {volume} {112}},\ \bibinfo {pages} {075301} (\bibinfo {year}
  {2014})}\BibitemShut {NoStop}%
\bibitem [{\citenamefont {Al~Khawaja}\ and\ \citenamefont
  {Stoof}(2001)}]{alkhawajaSkyrmionsFerromagneticBose2001a}%
  \BibitemOpen
  \bibfield  {author} {\bibinfo {author} {\bibfnamefont {Usama}\ \bibnamefont
  {Al~Khawaja}}\ and\ \bibinfo {author} {\bibfnamefont {Henk}\ \bibnamefont
  {Stoof}},\ }\bibfield  {title} {\enquote {\bibinfo {title} {Skyrmions in a
  ferromagnetic {Bose-Einstein} condensate},}\ }\href {\doibase
  10.1038/35082010} {\bibfield  {journal} {\bibinfo  {journal} {Nature
  (London)}\ }\textbf {\bibinfo {volume} {411}},\ \bibinfo {pages} {918--920}
  (\bibinfo {year} {2001})}\BibitemShut {NoStop}%
\bibitem [{\citenamefont {Kamihara}\ \emph {et~al.}(2006)\citenamefont
  {Kamihara}, \citenamefont {Hiramatsu}, \citenamefont {Hirano}, \citenamefont
  {Kawamura}, \citenamefont {Yanagi}, \citenamefont {Kamiya},\ and\
  \citenamefont {Hosono}}]{kamiharaIronBasedLayeredSuperconductor2006}%
  \BibitemOpen
  \bibfield  {author} {\bibinfo {author} {\bibfnamefont {Yoichi}\ \bibnamefont
  {Kamihara}}, \bibinfo {author} {\bibfnamefont {Hidenori}\ \bibnamefont
  {Hiramatsu}}, \bibinfo {author} {\bibfnamefont {Masahiro}\ \bibnamefont
  {Hirano}}, \bibinfo {author} {\bibfnamefont {Ryuto}\ \bibnamefont
  {Kawamura}}, \bibinfo {author} {\bibfnamefont {Hiroshi}\ \bibnamefont
  {Yanagi}}, \bibinfo {author} {\bibfnamefont {Toshio}\ \bibnamefont {Kamiya}},
  \ and\ \bibinfo {author} {\bibfnamefont {Hideo}\ \bibnamefont {Hosono}},\
  }\bibfield  {title} {\enquote {\bibinfo {title} {Iron-{{Based Layered
  Superconductor}}:\, {{LaOFeP}}},}\ }\href {\doibase 10.1021/ja063355c}
  {\bibfield  {journal} {\bibinfo  {journal} {J. Am. Chem. Soc.}\ }\textbf
  {\bibinfo {volume} {128}},\ \bibinfo {pages} {10012--10013} (\bibinfo {year}
  {2006})}\BibitemShut {NoStop}%
\bibitem [{\citenamefont {Kamihara}\ \emph {et~al.}(2008)\citenamefont
  {Kamihara}, \citenamefont {Watanabe}, \citenamefont {Hirano},\ and\
  \citenamefont {Hosono}}]{kamiharaIronBasedLayeredSuperconductor2008}%
  \BibitemOpen
  \bibfield  {author} {\bibinfo {author} {\bibfnamefont {Yoichi}\ \bibnamefont
  {Kamihara}}, \bibinfo {author} {\bibfnamefont {Takumi}\ \bibnamefont
  {Watanabe}}, \bibinfo {author} {\bibfnamefont {Masahiro}\ \bibnamefont
  {Hirano}}, \ and\ \bibinfo {author} {\bibfnamefont {Hideo}\ \bibnamefont
  {Hosono}},\ }\bibfield  {title} {\enquote {\bibinfo {title} {{Iron-Based
  Layered Superconductor La[O$_{1-x}$F$_x$]FeAs ($x = 0.05-0.12$) with $T_c =
  26~\mathrm{K}$}},}\ }\href {\doibase 10.1021/ja800073m} {\bibfield  {journal}
  {\bibinfo  {journal} {J. Am. Chem. Soc.}\ }\textbf {\bibinfo {volume}
  {130}},\ \bibinfo {pages} {3296--3297} (\bibinfo {year} {2008})}\BibitemShut
  {NoStop}%
\bibitem [{\citenamefont {Lin}\ \emph {et~al.}(2016)\citenamefont {Lin},
  \citenamefont {Maiti},\ and\ \citenamefont
  {Chubukov}}]{linDistinguishingIdPairing2016}%
  \BibitemOpen
  \bibfield  {author} {\bibinfo {author} {\bibfnamefont {Shi-Zeng}\
  \bibnamefont {Lin}}, \bibinfo {author} {\bibfnamefont {Saurabh}\ \bibnamefont
  {Maiti}}, \ and\ \bibinfo {author} {\bibfnamefont {Andrey}\ \bibnamefont
  {Chubukov}},\ }\bibfield  {title} {\enquote {\bibinfo {title} {Distinguishing
  between $s+id$ and $s+is$ pairing symmetries in multiband superconductors
  through spontaneous magnetization pattern induced by a defect},}\ }\href
  {\doibase 10.1103/PhysRevB.94.064519} {\bibfield  {journal} {\bibinfo
  {journal} {Phys. Rev. B}\ }\textbf {\bibinfo {volume} {94}},\ \bibinfo
  {pages} {064519} (\bibinfo {year} {2016})}\BibitemShut {NoStop}%
\bibitem [{\citenamefont {Garaud}\ \emph {et~al.}(2017)\citenamefont {Garaud},
  \citenamefont {Silaev},\ and\ \citenamefont
  {Babaev}}]{garaudMicroscopicallyDerivedMulticomponent2017}%
  \BibitemOpen
  \bibfield  {author} {\bibinfo {author} {\bibfnamefont {Julien}\ \bibnamefont
  {Garaud}}, \bibinfo {author} {\bibfnamefont {Mihail}\ \bibnamefont {Silaev}},
  \ and\ \bibinfo {author} {\bibfnamefont {Egor}\ \bibnamefont {Babaev}},\
  }\bibfield  {title} {\enquote {\bibinfo {title} {Microscopically derived
  multi-component {{Ginzburg}}\textendash{{Landau}} theories for $s+is$
  superconducting state},}\ }\href {\doibase 10.1016/j.physc.2016.07.010}
  {\bibfield  {journal} {\bibinfo  {journal} {Physica C}\ }\textbf {\bibinfo
  {volume} {533}},\ \bibinfo {pages} {63--73} (\bibinfo {year}
  {2017})}\BibitemShut {NoStop}%
\bibitem [{\citenamefont {Mazin}(2010)}]{mazinSuperconductivityGetsIron2010}%
  \BibitemOpen
  \bibfield  {author} {\bibinfo {author} {\bibfnamefont {Igor~I.}\ \bibnamefont
  {Mazin}},\ }\bibfield  {title} {\enquote {\bibinfo {title} {Superconductivity
  gets an iron boost},}\ }\href {\doibase 10.1038/nature08914} {\bibfield
  {journal} {\bibinfo  {journal} {Nature (London)}\ }\textbf {\bibinfo {volume}
  {464}},\ \bibinfo {pages} {183--186} (\bibinfo {year} {2010})}\BibitemShut
  {NoStop}%
\bibitem [{\citenamefont {Kuroki}\ \emph {et~al.}(2008)\citenamefont {Kuroki},
  \citenamefont {Onari}, \citenamefont {Arita}, \citenamefont {Usui},
  \citenamefont {Tanaka}, \citenamefont {Kontani},\ and\ \citenamefont
  {Aoki}}]{kurokiUnconventionalPairingOriginating2008}%
  \BibitemOpen
  \bibfield  {author} {\bibinfo {author} {\bibfnamefont {Kazuhiko}\
  \bibnamefont {Kuroki}}, \bibinfo {author} {\bibfnamefont {Seiichiro}\
  \bibnamefont {Onari}}, \bibinfo {author} {\bibfnamefont {Ryotaro}\
  \bibnamefont {Arita}}, \bibinfo {author} {\bibfnamefont {Hidetomo}\
  \bibnamefont {Usui}}, \bibinfo {author} {\bibfnamefont {Yukio}\ \bibnamefont
  {Tanaka}}, \bibinfo {author} {\bibfnamefont {Hiroshi}\ \bibnamefont
  {Kontani}}, \ and\ \bibinfo {author} {\bibfnamefont {Hideo}\ \bibnamefont
  {Aoki}},\ }\bibfield  {title} {\enquote {\bibinfo {title} {{Unconventional
  Pairing Originating from the Disconnected Fermi Surfaces of Superconducting
  ${\mathrm{LaFeAsO}}_{1\ensuremath{-}x}{\mathrm{F}}_{x}$}},}\ }\href {\doibase
  10.1103/PhysRevLett.101.087004} {\bibfield  {journal} {\bibinfo  {journal}
  {Phys. Rev. Lett.}\ }\textbf {\bibinfo {volume} {101}},\ \bibinfo {pages}
  {087004} (\bibinfo {year} {2008})}\BibitemShut {NoStop}%
\bibitem [{\citenamefont {Ding}\ \emph {et~al.}(2008)\citenamefont {Ding},
  \citenamefont {Richard}, \citenamefont {Nakayama}, \citenamefont {Sugawara},
  \citenamefont {Arakane}, \citenamefont {Sekiba}, \citenamefont {Takayama},
  \citenamefont {Souma}, \citenamefont {Sato}, \citenamefont {Takahashi},
  \citenamefont {Wang}, \citenamefont {Dai}, \citenamefont {Fang},
  \citenamefont {Chen}, \citenamefont {Luo},\ and\ \citenamefont
  {Wang}}]{dingObservationFermisurfaceDependent2008}%
  \BibitemOpen
  \bibfield  {author} {\bibinfo {author} {\bibfnamefont {H.}~\bibnamefont
  {Ding}}, \bibinfo {author} {\bibfnamefont {P.}~\bibnamefont {Richard}},
  \bibinfo {author} {\bibfnamefont {K.}~\bibnamefont {Nakayama}}, \bibinfo
  {author} {\bibfnamefont {K.}~\bibnamefont {Sugawara}}, \bibinfo {author}
  {\bibfnamefont {T.}~\bibnamefont {Arakane}}, \bibinfo {author} {\bibfnamefont
  {Y.}~\bibnamefont {Sekiba}}, \bibinfo {author} {\bibfnamefont
  {A.}~\bibnamefont {Takayama}}, \bibinfo {author} {\bibfnamefont
  {S.}~\bibnamefont {Souma}}, \bibinfo {author} {\bibfnamefont
  {T.}~\bibnamefont {Sato}}, \bibinfo {author} {\bibfnamefont {T.}~\bibnamefont
  {Takahashi}}, \bibinfo {author} {\bibfnamefont {Z.}~\bibnamefont {Wang}},
  \bibinfo {author} {\bibfnamefont {X.}~\bibnamefont {Dai}}, \bibinfo {author}
  {\bibfnamefont {Z.}~\bibnamefont {Fang}}, \bibinfo {author} {\bibfnamefont
  {G.~F.}\ \bibnamefont {Chen}}, \bibinfo {author} {\bibfnamefont {J.~L.}\
  \bibnamefont {Luo}}, \ and\ \bibinfo {author} {\bibfnamefont {N.~L.}\
  \bibnamefont {Wang}},\ }\bibfield  {title} {\enquote {\bibinfo {title}
  {Observation of {{Fermi}}-surface-dependent nodeless superconducting gaps in
  {${\text{Ba}}_{0.6}{\text{K}}_{0.4}{\text{Fe}}_{2}{\text{As}}_{2}$}},}\
  }\href {\doibase 10.1209/0295-5075/83/47001} {\bibfield  {journal} {\bibinfo
  {journal} {EPL}\ }\textbf {\bibinfo {volume} {83}},\ \bibinfo {pages} {47001}
  (\bibinfo {year} {2008})}\BibitemShut {NoStop}%
\bibitem [{\citenamefont {Luo}\ \emph {et~al.}(2009)\citenamefont {Luo},
  \citenamefont {Tanatar}, \citenamefont {Reid}, \citenamefont {Shakeripour},
  \citenamefont {Doiron-Leyraud}, \citenamefont {Ni}, \citenamefont {Bud'ko},
  \citenamefont {Canfield}, \citenamefont {Luo}, \citenamefont {Wang},
  \citenamefont {Wen}, \citenamefont {Prozorov},\ and\ \citenamefont
  {Taillefer}}]{luoQuasiparticleHeatTransport2009}%
  \BibitemOpen
  \bibfield  {author} {\bibinfo {author} {\bibfnamefont {X.~G.}\ \bibnamefont
  {Luo}}, \bibinfo {author} {\bibfnamefont {M.~A.}\ \bibnamefont {Tanatar}},
  \bibinfo {author} {\bibfnamefont {J.-Ph.}\ \bibnamefont {Reid}}, \bibinfo
  {author} {\bibfnamefont {H.}~\bibnamefont {Shakeripour}}, \bibinfo {author}
  {\bibfnamefont {N.}~\bibnamefont {Doiron-Leyraud}}, \bibinfo {author}
  {\bibfnamefont {N.}~\bibnamefont {Ni}}, \bibinfo {author} {\bibfnamefont
  {S.~L.}\ \bibnamefont {Bud'ko}}, \bibinfo {author} {\bibfnamefont {P.~C.}\
  \bibnamefont {Canfield}}, \bibinfo {author} {\bibfnamefont {Huiqian}\
  \bibnamefont {Luo}}, \bibinfo {author} {\bibfnamefont {Zhaosheng}\
  \bibnamefont {Wang}}, \bibinfo {author} {\bibfnamefont {Hai-Hu}\ \bibnamefont
  {Wen}}, \bibinfo {author} {\bibfnamefont {R.}~\bibnamefont {Prozorov}}, \
  and\ \bibinfo {author} {\bibfnamefont {Louis}\ \bibnamefont {Taillefer}},\
  }\bibfield  {title} {\enquote {\bibinfo {title} {Quasiparticle heat transport
  in single-crystalline
  {${\text{Ba}}_{1\ensuremath{-}x}{\text{K}}_{x}{\text{Fe}}_{2}{\text{As}}_{2}$}:
  Evidence for a $k$-dependent superconducting gap without nodes},}\ }\href
  {\doibase 10.1103/PhysRevB.80.140503} {\bibfield  {journal} {\bibinfo
  {journal} {Phys. Rev. B}\ }\textbf {\bibinfo {volume} {80}},\ \bibinfo
  {pages} {140503(R)} (\bibinfo {year} {2009})}\BibitemShut {NoStop}%
\bibitem [{\citenamefont {Christianson}\ \emph {et~al.}(2008)\citenamefont
  {Christianson}, \citenamefont {Goremychkin}, \citenamefont {Osborn},
  \citenamefont {Rosenkranz}, \citenamefont {Lumsden}, \citenamefont
  {Malliakas}, \citenamefont {Todorov}, \citenamefont {Claus}, \citenamefont
  {Chung}, \citenamefont {Kanatzidis}, \citenamefont {Bewley},\ and\
  \citenamefont {Guidi}}]{christiansonUnconventionalSuperconductivityBa02008}%
  \BibitemOpen
  \bibfield  {author} {\bibinfo {author} {\bibfnamefont {A.~D.}\ \bibnamefont
  {Christianson}}, \bibinfo {author} {\bibfnamefont {E.~A.}\ \bibnamefont
  {Goremychkin}}, \bibinfo {author} {\bibfnamefont {R.}~\bibnamefont {Osborn}},
  \bibinfo {author} {\bibfnamefont {S.}~\bibnamefont {Rosenkranz}}, \bibinfo
  {author} {\bibfnamefont {M.~D.}\ \bibnamefont {Lumsden}}, \bibinfo {author}
  {\bibfnamefont {C.~D.}\ \bibnamefont {Malliakas}}, \bibinfo {author}
  {\bibfnamefont {I.~S.}\ \bibnamefont {Todorov}}, \bibinfo {author}
  {\bibfnamefont {H.}~\bibnamefont {Claus}}, \bibinfo {author} {\bibfnamefont
  {D.~Y.}\ \bibnamefont {Chung}}, \bibinfo {author} {\bibfnamefont {M.~G.}\
  \bibnamefont {Kanatzidis}}, \bibinfo {author} {\bibfnamefont {R.~I.}\
  \bibnamefont {Bewley}}, \ and\ \bibinfo {author} {\bibfnamefont
  {T.}~\bibnamefont {Guidi}},\ }\bibfield  {title} {\enquote {\bibinfo {title}
  {Unconventional superconductivity in
  {{Ba}}{\textsubscript{0.6}}{{K}}{\textsubscript{0.4}}{{Fe}}{\textsubscript{2}}{{As}}{\textsubscript{2}}
  from inelastic neutron scattering},}\ }\href {\doibase 10.1038/nature07625}
  {\bibfield  {journal} {\bibinfo  {journal} {Nature (London)}\ }\textbf
  {\bibinfo {volume} {456}},\ \bibinfo {pages} {930--932} (\bibinfo {year}
  {2008})}\BibitemShut {NoStop}%
\bibitem [{\citenamefont {Xu}\ \emph {et~al.}(2013)\citenamefont {Xu},
  \citenamefont {Richard}, \citenamefont {Shi}, \citenamefont {van Roekeghem},
  \citenamefont {Qian}, \citenamefont {Razzoli}, \citenamefont {Rienks},
  \citenamefont {Chen}, \citenamefont {Ieki}, \citenamefont {Nakayama},
  \citenamefont {Sato}, \citenamefont {Takahashi}, \citenamefont {Shi},\ and\
  \citenamefont {Ding}}]{xuPossibleNodalSuperconducting2013}%
  \BibitemOpen
  \bibfield  {author} {\bibinfo {author} {\bibfnamefont {N.}~\bibnamefont
  {Xu}}, \bibinfo {author} {\bibfnamefont {P.}~\bibnamefont {Richard}},
  \bibinfo {author} {\bibfnamefont {X.}~\bibnamefont {Shi}}, \bibinfo {author}
  {\bibfnamefont {A.}~\bibnamefont {van Roekeghem}}, \bibinfo {author}
  {\bibfnamefont {T.}~\bibnamefont {Qian}}, \bibinfo {author} {\bibfnamefont
  {E.}~\bibnamefont {Razzoli}}, \bibinfo {author} {\bibfnamefont
  {E.}~\bibnamefont {Rienks}}, \bibinfo {author} {\bibfnamefont {G.-F.}\
  \bibnamefont {Chen}}, \bibinfo {author} {\bibfnamefont {E.}~\bibnamefont
  {Ieki}}, \bibinfo {author} {\bibfnamefont {K.}~\bibnamefont {Nakayama}},
  \bibinfo {author} {\bibfnamefont {T.}~\bibnamefont {Sato}}, \bibinfo {author}
  {\bibfnamefont {T.}~\bibnamefont {Takahashi}}, \bibinfo {author}
  {\bibfnamefont {M.}~\bibnamefont {Shi}}, \ and\ \bibinfo {author}
  {\bibfnamefont {H.}~\bibnamefont {Ding}},\ }\bibfield  {title} {\enquote
  {\bibinfo {title} {Possible nodal superconducting gap and lifshitz transition
  in heavily hole-doped
  {${\text{Ba}}_{0.1}{\text{K}}_{0.9}{\text{Fe}}_{2}{\text{As}}_{2}$}},}\
  }\href {\doibase 10.1103/PhysRevB.88.220508} {\bibfield  {journal} {\bibinfo
  {journal} {Phys. Rev. B}\ }\textbf {\bibinfo {volume} {88}},\ \bibinfo
  {pages} {220508(R)} (\bibinfo {year} {2013})}\BibitemShut {NoStop}%
\bibitem [{\citenamefont {Maiti}\ and\ \citenamefont
  {Chubukov}(2013)}]{maitiStateBrokenTimereversal2013}%
  \BibitemOpen
  \bibfield  {author} {\bibinfo {author} {\bibfnamefont {Saurabh}\ \bibnamefont
  {Maiti}}\ and\ \bibinfo {author} {\bibfnamefont {Andrey~V.}\ \bibnamefont
  {Chubukov}},\ }\bibfield  {title} {\enquote {\bibinfo {title} {$s+is$ state
  with broken time-reversal symmetry in {Fe}-based superconductors},}\ }\href
  {\doibase 10.1103/PhysRevB.87.144511} {\bibfield  {journal} {\bibinfo
  {journal} {Phys. Rev. B}\ }\textbf {\bibinfo {volume} {87}},\ \bibinfo
  {pages} {144511} (\bibinfo {year} {2013})}\BibitemShut {NoStop}%
\bibitem [{\citenamefont {Maiti}\ \emph {et~al.}(2015)\citenamefont {Maiti},
  \citenamefont {Sigrist},\ and\ \citenamefont
  {Chubukov}}]{maitiSpontaneousCurrentsSuperconductor2015}%
  \BibitemOpen
  \bibfield  {author} {\bibinfo {author} {\bibfnamefont {Saurabh}\ \bibnamefont
  {Maiti}}, \bibinfo {author} {\bibfnamefont {Manfred}\ \bibnamefont
  {Sigrist}}, \ and\ \bibinfo {author} {\bibfnamefont {Andrey}\ \bibnamefont
  {Chubukov}},\ }\bibfield  {title} {\enquote {\bibinfo {title} {Spontaneous
  currents in a superconductor with $s+is$ symmetry},}\ }\href {\doibase
  10.1103/PhysRevB.91.161102} {\bibfield  {journal} {\bibinfo  {journal} {Phys.
  Rev. B}\ }\textbf {\bibinfo {volume} {91}},\ \bibinfo {pages} {161102(R)}
  (\bibinfo {year} {2015})}\BibitemShut {NoStop}%
\bibitem [{\citenamefont {Lee}\ \emph {et~al.}(2009)\citenamefont {Lee},
  \citenamefont {Zhang},\ and\ \citenamefont
  {Wu}}]{leePairingStateTimeReversal2009}%
  \BibitemOpen
  \bibfield  {author} {\bibinfo {author} {\bibfnamefont {Wei-Cheng}\
  \bibnamefont {Lee}}, \bibinfo {author} {\bibfnamefont {Shou-Cheng}\
  \bibnamefont {Zhang}}, \ and\ \bibinfo {author} {\bibfnamefont {Congjun}\
  \bibnamefont {Wu}},\ }\bibfield  {title} {\enquote {\bibinfo {title} {Pairing
  state with a time-reversal symmetry breaking in {FeAs}-based
  superconductors},}\ }\href {\doibase 10.1103/PhysRevLett.102.217002}
  {\bibfield  {journal} {\bibinfo  {journal} {Phys. Rev. Lett.}\ }\textbf
  {\bibinfo {volume} {102}},\ \bibinfo {pages} {217002} (\bibinfo {year}
  {2009})}\BibitemShut {NoStop}%
\bibitem [{\citenamefont {Grinenko}\ \emph {et~al.}(2017)\citenamefont
  {Grinenko}, \citenamefont {Materne}, \citenamefont {Sarkar}, \citenamefont
  {Luetkens}, \citenamefont {Kihou}, \citenamefont {Lee}, \citenamefont
  {Akhmadaliev}, \citenamefont {Efremov}, \citenamefont {Drechsler},\ and\
  \citenamefont {Klauss}}]{grinenkoSuperconductivityBrokenTimereversal2017}%
  \BibitemOpen
  \bibfield  {author} {\bibinfo {author} {\bibfnamefont {V.}~\bibnamefont
  {Grinenko}}, \bibinfo {author} {\bibfnamefont {P.}~\bibnamefont {Materne}},
  \bibinfo {author} {\bibfnamefont {R.}~\bibnamefont {Sarkar}}, \bibinfo
  {author} {\bibfnamefont {H.}~\bibnamefont {Luetkens}}, \bibinfo {author}
  {\bibfnamefont {K.}~\bibnamefont {Kihou}}, \bibinfo {author} {\bibfnamefont
  {C.~H.}\ \bibnamefont {Lee}}, \bibinfo {author} {\bibfnamefont
  {S.}~\bibnamefont {Akhmadaliev}}, \bibinfo {author} {\bibfnamefont {D.~V.}\
  \bibnamefont {Efremov}}, \bibinfo {author} {\bibfnamefont {S.-L.}\
  \bibnamefont {Drechsler}}, \ and\ \bibinfo {author} {\bibfnamefont {H.-H.}\
  \bibnamefont {Klauss}},\ }\bibfield  {title} {\enquote {\bibinfo {title}
  {Superconductivity with broken time-reversal symmetry in ion-irradiated
  {${\mathbf{Ba}}_{\mathbf{0}.\mathbf{27}}{\mathbf{K}}_{\mathbf{0}.\mathbf{73}}{\mathbf{Fe}}_{\mathbf{2}}{\mathbf{As}}_{\mathbf{2}}$}
  single crystals},}\ }\href {\doibase 10.1103/PhysRevB.95.214511} {\bibfield
  {journal} {\bibinfo  {journal} {Phys. Rev. B}\ }\textbf {\bibinfo {volume}
  {95}},\ \bibinfo {pages} {214511} (\bibinfo {year} {2017})}\BibitemShut
  {NoStop}%
\bibitem [{\citenamefont {Silaev}\ \emph {et~al.}(2015)\citenamefont {Silaev},
  \citenamefont {Garaud},\ and\ \citenamefont
  {Babaev}}]{silaevUnconventionalThermoelectricEffect2015}%
  \BibitemOpen
  \bibfield  {author} {\bibinfo {author} {\bibfnamefont {Mihail}\ \bibnamefont
  {Silaev}}, \bibinfo {author} {\bibfnamefont {Julien}\ \bibnamefont {Garaud}},
  \ and\ \bibinfo {author} {\bibfnamefont {Egor}\ \bibnamefont {Babaev}},\
  }\bibfield  {title} {\enquote {\bibinfo {title} {Unconventional
  thermoelectric effect in superconductors that break time-reversal
  symmetry},}\ }\href {\doibase 10.1103/PhysRevB.92.174510} {\bibfield
  {journal} {\bibinfo  {journal} {Phys. Rev. B}\ }\textbf {\bibinfo {volume}
  {92}},\ \bibinfo {pages} {174510} (\bibinfo {year} {2015})}\BibitemShut
  {NoStop}%
\bibitem [{\citenamefont {Garaud}\ \emph
  {et~al.}(2016{\natexlab{b}})\citenamefont {Garaud}, \citenamefont {Silaev},\
  and\ \citenamefont
  {Babaev}}]{garaudThermoelectricSignaturesTimeReversal2016}%
  \BibitemOpen
  \bibfield  {author} {\bibinfo {author} {\bibfnamefont {Julien}\ \bibnamefont
  {Garaud}}, \bibinfo {author} {\bibfnamefont {Mihail}\ \bibnamefont {Silaev}},
  \ and\ \bibinfo {author} {\bibfnamefont {Egor}\ \bibnamefont {Babaev}},\
  }\bibfield  {title} {\enquote {\bibinfo {title} {Thermoelectric signatures of
  time-reversal symmetry breaking states in multiband superconductors},}\
  }\href {\doibase 10.1103/PhysRevLett.116.097002} {\bibfield  {journal}
  {\bibinfo  {journal} {Phys. Rev. Lett.}\ }\textbf {\bibinfo {volume} {116}},\
  \bibinfo {pages} {097002} (\bibinfo {year} {2016}{\natexlab{b}})}\BibitemShut
  {NoStop}%
\bibitem [{\citenamefont {Ren}\ \emph {et~al.}(1995)\citenamefont {Ren},
  \citenamefont {Xu},\ and\ \citenamefont
  {Ting}}]{renGinzburgLandauEquationsVortex1995}%
  \BibitemOpen
  \bibfield  {author} {\bibinfo {author} {\bibfnamefont {Yong}\ \bibnamefont
  {Ren}}, \bibinfo {author} {\bibfnamefont {Ji-Hai}\ \bibnamefont {Xu}}, \ and\
  \bibinfo {author} {\bibfnamefont {C.~S.}\ \bibnamefont {Ting}},\ }\bibfield
  {title} {\enquote {\bibinfo {title} {Ginzburg-landau equations and vortex
  structure of a ${d}_{{x}^{2}{\ensuremath{-}y}^{2}}$ superconductor},}\ }\href
  {\doibase 10.1103/PhysRevLett.74.3680} {\bibfield  {journal} {\bibinfo
  {journal} {Phys. Rev. Lett.}\ }\textbf {\bibinfo {volume} {74}},\ \bibinfo
  {pages} {3680--3683} (\bibinfo {year} {1995})}\BibitemShut {NoStop}%
\bibitem [{\citenamefont {Xu}\ \emph {et~al.}(1996)\citenamefont {Xu},
  \citenamefont {Ren},\ and\ \citenamefont
  {Ting}}]{xuStructuresSingleVortex1996}%
  \BibitemOpen
  \bibfield  {author} {\bibinfo {author} {\bibfnamefont {Ji-Hai}\ \bibnamefont
  {Xu}}, \bibinfo {author} {\bibfnamefont {Yong}\ \bibnamefont {Ren}}, \ and\
  \bibinfo {author} {\bibfnamefont {Chin-Sen}\ \bibnamefont {Ting}},\
  }\bibfield  {title} {\enquote {\bibinfo {title} {Structures of single vortex
  and vortex lattice in a $d$-wave superconductor},}\ }\href {\doibase
  10.1103/PhysRevB.53.R2991} {\bibfield  {journal} {\bibinfo  {journal} {Phys.
  Rev. B}\ }\textbf {\bibinfo {volume} {53}},\ \bibinfo {pages} {R2991--R2994}
  (\bibinfo {year} {1996})}\BibitemShut {NoStop}%
\bibitem [{\citenamefont {Franz}\ \emph {et~al.}(1996)\citenamefont {Franz},
  \citenamefont {Kallin}, \citenamefont {Soininen}, \citenamefont {Berlinsky},\
  and\ \citenamefont {Fetter}}]{franzVortexStateDwave1996}%
  \BibitemOpen
  \bibfield  {author} {\bibinfo {author} {\bibfnamefont {M.}~\bibnamefont
  {Franz}}, \bibinfo {author} {\bibfnamefont {C.}~\bibnamefont {Kallin}},
  \bibinfo {author} {\bibfnamefont {P.~I.}\ \bibnamefont {Soininen}}, \bibinfo
  {author} {\bibfnamefont {A.~J.}\ \bibnamefont {Berlinsky}}, \ and\ \bibinfo
  {author} {\bibfnamefont {A.~L.}\ \bibnamefont {Fetter}},\ }\bibfield  {title}
  {\enquote {\bibinfo {title} {Vortex state in a d-wave superconductor},}\
  }\href {\doibase 10.1103/PhysRevB.53.5795} {\bibfield  {journal} {\bibinfo
  {journal} {Phys. Rev. B}\ }\textbf {\bibinfo {volume} {53}},\ \bibinfo
  {pages} {5795--5814} (\bibinfo {year} {1996})}\BibitemShut {NoStop}%
\bibitem [{\citenamefont {Li}\ \emph {et~al.}(1999)\citenamefont {Li},
  \citenamefont {Wang},\ and\ \citenamefont
  {Wang}}]{liVortexStructureWave1999}%
  \BibitemOpen
  \bibfield  {author} {\bibinfo {author} {\bibfnamefont {Qunqing}\ \bibnamefont
  {Li}}, \bibinfo {author} {\bibfnamefont {Z.~D.}\ \bibnamefont {Wang}}, \ and\
  \bibinfo {author} {\bibfnamefont {Qiang-Hua}\ \bibnamefont {Wang}},\
  }\bibfield  {title} {\enquote {\bibinfo {title} {Vortex structure for a
  $d+is$-wave superconductor},}\ }\href {\doibase 10.1103/PhysRevB.59.613}
  {\bibfield  {journal} {\bibinfo  {journal} {Phys. Rev. B}\ }\textbf {\bibinfo
  {volume} {59}},\ \bibinfo {pages} {613--618} (\bibinfo {year}
  {1999})}\BibitemShut {NoStop}%
\bibitem [{\citenamefont
  {Volovik}(1993)}]{volovikSuperconductivityLinesGAP1993}%
  \BibitemOpen
  \bibfield  {author} {\bibinfo {author} {\bibfnamefont {G.~E.}\ \bibnamefont
  {Volovik}},\ }\bibfield  {title} {\enquote {\bibinfo {title}
  {Superconductivity with lines of gap nodes: Density of states in the
  vortex},}\ }\href@noop {} {\bibfield  {journal} {\bibinfo  {journal} {JETP
  Lett.}\ }\textbf {\bibinfo {volume} {58}},\ \bibinfo {pages} {469} (\bibinfo
  {year} {1993})}\BibitemShut {NoStop}%
\bibitem [{\citenamefont {Du}\ \emph {et~al.}(1993)\citenamefont {Du},
  \citenamefont {Gunburger},\ and\ \citenamefont
  {Peterson}}]{duModelingAnalysisPeriodic1993}%
  \BibitemOpen
  \bibfield  {author} {\bibinfo {author} {\bibfnamefont {Q.}~\bibnamefont
  {Du}}, \bibinfo {author} {\bibfnamefont {M.}~\bibnamefont {Gunburger}}, \
  and\ \bibinfo {author} {\bibfnamefont {J.}~\bibnamefont {Peterson}},\
  }\bibfield  {title} {\enquote {\bibinfo {title} {Modeling and {{Analysis}} of
  a {{Periodic Ginzburg}}\textendash{{Landau Model}} for {{Type}}-{{II
  Superconductors}}},}\ }\href {\doibase 10.1137/0153035} {\bibfield  {journal}
  {\bibinfo  {journal} {SIAM J. Appl. Math.}\ }\textbf {\bibinfo {volume}
  {53}},\ \bibinfo {pages} {689--717} (\bibinfo {year} {1993})}\BibitemShut
  {NoStop}%
\bibitem [{\citenamefont {Garaud}\ and\ \citenamefont
  {Babaev}(2015{\natexlab{b}})}]{garaudVortexChainsDue2015}%
  \BibitemOpen
  \bibfield  {author} {\bibinfo {author} {\bibfnamefont {Julien}\ \bibnamefont
  {Garaud}}\ and\ \bibinfo {author} {\bibfnamefont {Egor}\ \bibnamefont
  {Babaev}},\ }\bibfield  {title} {\enquote {\bibinfo {title} {Vortex chains
  due to nonpairwise interactions and field-induced phase transitions between
  states with different broken symmetry in superconductors with competing order
  parameters},}\ }\href {\doibase 10.1103/PhysRevB.91.014510} {\bibfield
  {journal} {\bibinfo  {journal} {Phys. Rev. B}\ }\textbf {\bibinfo {volume}
  {91}},\ \bibinfo {pages} {014510} (\bibinfo {year}
  {2015}{\natexlab{b}})}\BibitemShut {NoStop}%
\bibitem [{\citenamefont {Shen}\ and\ \citenamefont
  {Davis}(2008)}]{shenCuprateHighTcSuperconductors2008}%
  \BibitemOpen
  \bibfield  {author} {\bibinfo {author} {\bibfnamefont {Kyle~M.}\ \bibnamefont
  {Shen}}\ and\ \bibinfo {author} {\bibfnamefont {J.~C.~Seamus}\ \bibnamefont
  {Davis}},\ }\bibfield  {title} {\enquote {\bibinfo {title} {Cuprate
  high-{{Tc}} superconductors},}\ }\href {\doibase
  10.1016/S1369-7021(08)70175-5} {\bibfield  {journal} {\bibinfo  {journal}
  {Materials Today}\ }\textbf {\bibinfo {volume} {11}},\ \bibinfo {pages}
  {14--21} (\bibinfo {year} {2008})}\BibitemShut {NoStop}%
\bibitem [{\citenamefont {Zhong}\ \emph {et~al.}(2016)\citenamefont {Zhong},
  \citenamefont {Wang}, \citenamefont {Han}, \citenamefont {Lv}, \citenamefont
  {Wang}, \citenamefont {Zhang}, \citenamefont {Ding}, \citenamefont {Zhang},
  \citenamefont {Wang}, \citenamefont {He}, \citenamefont {Zhong},
  \citenamefont {Schneeloch}, \citenamefont {Gu}, \citenamefont {Song},
  \citenamefont {Ma},\ and\ \citenamefont
  {Xue}}]{zhongNodelessPairingSuperconducting2016}%
  \BibitemOpen
  \bibfield  {author} {\bibinfo {author} {\bibfnamefont {Yong}\ \bibnamefont
  {Zhong}}, \bibinfo {author} {\bibfnamefont {Yang}\ \bibnamefont {Wang}},
  \bibinfo {author} {\bibfnamefont {Sha}\ \bibnamefont {Han}}, \bibinfo
  {author} {\bibfnamefont {Yan-Feng}\ \bibnamefont {Lv}}, \bibinfo {author}
  {\bibfnamefont {Wen-Lin}\ \bibnamefont {Wang}}, \bibinfo {author}
  {\bibfnamefont {Ding}\ \bibnamefont {Zhang}}, \bibinfo {author}
  {\bibfnamefont {Hao}\ \bibnamefont {Ding}}, \bibinfo {author} {\bibfnamefont
  {Yi-Min}\ \bibnamefont {Zhang}}, \bibinfo {author} {\bibfnamefont {Lili}\
  \bibnamefont {Wang}}, \bibinfo {author} {\bibfnamefont {Ke}~\bibnamefont
  {He}}, \bibinfo {author} {\bibfnamefont {Ruidan}\ \bibnamefont {Zhong}},
  \bibinfo {author} {\bibfnamefont {John~A.}\ \bibnamefont {Schneeloch}},
  \bibinfo {author} {\bibfnamefont {Gen-Da}\ \bibnamefont {Gu}}, \bibinfo
  {author} {\bibfnamefont {Can-Li}\ \bibnamefont {Song}}, \bibinfo {author}
  {\bibfnamefont {Xu-Cun}\ \bibnamefont {Ma}}, \ and\ \bibinfo {author}
  {\bibfnamefont {Qi-Kun}\ \bibnamefont {Xue}},\ }\bibfield  {title} {\enquote
  {\bibinfo {title} {Nodeless pairing in superconducting copper-oxide monolayer
  films on {Bi$_2$Sr$_2$CaCu$_2$O$_{8+\delta}$}},}\ }\href {\doibase
  10.1007/s11434-016-1145-4} {\bibfield  {journal} {\bibinfo  {journal} {Sci.
  Bull.}\ }\textbf {\bibinfo {volume} {61}},\ \bibinfo {pages} {1239--1247}
  (\bibinfo {year} {2016})}\BibitemShut {NoStop}%
\bibitem [{\citenamefont {Jiang}\ \emph {et~al.}(2018)\citenamefont {Jiang},
  \citenamefont {Wu}, \citenamefont {Hu},\ and\ \citenamefont
  {Wang}}]{jiangNodelessHighSuperconductivity2018}%
  \BibitemOpen
  \bibfield  {author} {\bibinfo {author} {\bibfnamefont {Kun}\ \bibnamefont
  {Jiang}}, \bibinfo {author} {\bibfnamefont {Xianxin}\ \bibnamefont {Wu}},
  \bibinfo {author} {\bibfnamefont {Jiangping}\ \bibnamefont {Hu}}, \ and\
  \bibinfo {author} {\bibfnamefont {Ziqiang}\ \bibnamefont {Wang}},\ }\bibfield
   {title} {\enquote {\bibinfo {title} {Nodeless high-${T}_{c}$
  superconductivity in the highly overdoped {${\mathrm{CuO}}_{2}$}
  monolayer},}\ }\href {\doibase 10.1103/PhysRevLett.121.227002} {\bibfield
  {journal} {\bibinfo  {journal} {Phys. Rev. Lett.}\ }\textbf {\bibinfo
  {volume} {121}},\ \bibinfo {pages} {227002} (\bibinfo {year}
  {2018})}\BibitemShut {NoStop}%
\bibitem [{\citenamefont {Zhu}(2019)}]{PhysRevLett.122.236401}%
  \BibitemOpen
  \bibfield  {author} {\bibinfo {author} {\bibfnamefont {Xiaoyu}\ \bibnamefont
  {Zhu}},\ }\bibfield  {title} {\enquote {\bibinfo {title} {Second-order
  topological superconductors with mixed pairing},}\ }\href {\doibase
  10.1103/PhysRevLett.122.236401} {\bibfield  {journal} {\bibinfo  {journal}
  {Phys. Rev. Lett.}\ }\textbf {\bibinfo {volume} {122}},\ \bibinfo {pages}
  {236401} (\bibinfo {year} {2019})}\BibitemShut {NoStop}%
\bibitem [{\citenamefont {Hu}\ \emph {et~al.}(1997)\citenamefont {Hu},
  \citenamefont {Miyashita},\ and\ \citenamefont
  {Tachiki}}]{huFunctionPeakSpecific1997}%
  \BibitemOpen
  \bibfield  {author} {\bibinfo {author} {\bibfnamefont {Xiao}\ \bibnamefont
  {Hu}}, \bibinfo {author} {\bibfnamefont {Seiji}\ \bibnamefont {Miyashita}}, \
  and\ \bibinfo {author} {\bibfnamefont {Masashi}\ \bibnamefont {Tachiki}},\
  }\bibfield  {title} {\enquote {\bibinfo {title}
  {$\mathit{\ensuremath{\delta}}$-function peak in the specific heat of high-
  ${T}_{c}$ superconductors: Monte carlo simulation},}\ }\href {\doibase
  10.1103/PhysRevLett.79.3498} {\bibfield  {journal} {\bibinfo  {journal}
  {Phys. Rev. Lett.}\ }\textbf {\bibinfo {volume} {79}},\ \bibinfo {pages}
  {3498--3501} (\bibinfo {year} {1997})}\BibitemShut {NoStop}%
\bibitem [{\citenamefont {Sm\o{}rgrav}\ \emph {et~al.}(2005)\citenamefont
  {Sm\o{}rgrav}, \citenamefont {Smiseth}, \citenamefont {Babaev},\ and\
  \citenamefont {Sudb\o{}}}]{smorgravVortexSublatticeMelting2005}%
  \BibitemOpen
  \bibfield  {author} {\bibinfo {author} {\bibfnamefont {E.}~\bibnamefont
  {Sm\o{}rgrav}}, \bibinfo {author} {\bibfnamefont {J.}~\bibnamefont
  {Smiseth}}, \bibinfo {author} {\bibfnamefont {E.}~\bibnamefont {Babaev}}, \
  and\ \bibinfo {author} {\bibfnamefont {A.}~\bibnamefont {Sudb\o{}}},\
  }\bibfield  {title} {\enquote {\bibinfo {title} {Vortex sublattice melting in
  a two-component superconductor},}\ }\href {\doibase
  10.1103/PhysRevLett.94.096401} {\bibfield  {journal} {\bibinfo  {journal}
  {Phys. Rev. Lett.}\ }\textbf {\bibinfo {volume} {94}},\ \bibinfo {pages}
  {096401} (\bibinfo {year} {2005})}\BibitemShut {NoStop}%
\end{thebibliography}

%

\end{document}